\input harvmac
\newcount\figno
\figno=0
\def\fig#1#2#3{
\par\begingroup\parindent=0pt\leftskip=1cm\rightskip=1cm\parindent=0pt
\global\advance\figno by 1
\midinsert
\epsfxsize=#3
\centerline{\epsfbox{#2}}
\vskip 12pt
{\bf Fig. \the\figno:} #1\par
\endinsert\endgroup\par
}
\def\figlabel#1{\xdef#1{\the\figno}}
\def\encadremath#1{\vbox{\hrule\hbox{\vrule\kern8pt\vbox{\kern8pt
\hbox{$\displaystyle #1$}\kern8pt}
\kern8pt\vrule}\hrule}}

\overfullrule=0pt
\def\Spin{{\rm Spin}}
%macros
%
\def\tilde{\widetilde}
\def\bar{\overline}
\def\Z{{\bf Z}}
\def\T{{\bf T}}
\def\S{{\bf S}}
\def\R{{\bf R}}

\font\zfont = cmss10 %scaled \magstep1
\font\litfont = cmr6

\def\bigone{\hbox{1\kern -.23em {\rm l}}}
\def\ZZ{\hbox{\zfont Z\kern-.4emZ}}
\def\half{{\litfont {1 \over 2}}}

\def\CM{{{\cal {M}}}}

\Title{hep-th/9712028, IASSNS-HEP-97-129}
{\vbox{\centerline{TOROIDAL COMPACTIFICATION }
\bigskip
\centerline{ WITHOUT VECTOR STRUCTURE}}}
\smallskip
\centerline{Edward Witten}
\smallskip
\centerline{\it School of Natural Sciences, Institute for Advanced Study}
\centerline{\it Olden Lane, Princeton, NJ 08540, USA}\bigskip

\medskip

\noindent

Many important ideas about string duality that appear in 
conventional $\T^2$ compactification have analogs for $\T^2$
compactification without vector structure.
We analyze some of these issues and show, in particular, how
orientifold planes associated with $Sp(n)$ gauge groups can
arise from $T$-duality and how they can be interpreted in $F$-theory.
We also, in an appendix, resolve a longstanding puzzle concerning
the computation of $\Tr\,(-1)^F$ in four-dimensional supersymmetric
Yang-Mills theory with gauge group $SO(n)$.

\Date{December, 1997}
%text of paper
\newsec{Introduction}

The gauge group of what is often informally called the $SO(32)$
heterotic string is actually $\Spin(32)/\Z_2$, where the $\Z_2$ is generated
by an element $x$ of $\Spin(32)$ which if projected to $SO(32)$  
becomes the generator $-1$ of the center of $SO(32)$ 
\ref\ghmr{D. J. Gross, J. A. Harvey, E. Martinec, and R. Rohm,
``Heterotic String Theory (I): The Free Heterotic String,''
Nucl. Phys. {\bf B256} (1985) 253.}.
In particular, $x$ acts as $-1$ in the 32 dimensional representation
of $SO(32)$, so this representation is not present for the heterotic
string.  Likewise, $x$ acts as $\pm 1$ for positive and negative chirality
spinors of $\Spin(32)$, so only the positive chirality spinors are present.

Because particles in the vector (or negative chirality
spinor) representation of $\Spin(32)$ are absent in the theory, it is
possible to consider compactifications of the heterotic string in which
the topology of the gauge bundle is such that these representations
would actually be impossible.  This happens if Dirac quantization
is obeyed for $\Spin(32)/\Z_2$ representations, but not for the vector
representation of $SO(32)$.

Given a $\Spin(32)/\Z_2$ bundle $V$ over a spacetime manifold $M$, the
obstruction to ``vector structure'' is measured by a mod two
cohomology class $\tilde w_2(M)$, which assigns the value $+1$ to any
two-cycle in $M$ on which there is vector structure, and $-1$ for
those for which there is not. The name $\tilde w_2$ is intended
to reflect the analogy with the second Stieffel-Whitney class
$w_2$, which is the obstruction to spin structure.\foot{A formal
definition of $\tilde w_2(V)$ is as follows.  Cover $M$ with small
open sets $U_i$ on which $V$ is trivial. Let $V_{ij}$ be 
$\Spin(32)/\Z_2$-valued
transition functions on $U_i\cap U_j$, defining $V$.  So in particular
$V_{ij}V_{jl}V_{li}=1$ in $U_i\cap U_j\cap U_l$ for all $i,j,$ and $l$.
Pick a set
of liftings $\tilde V_{ij}$ of $V_{ij}$ to $\Spin(32)$.  Then
$W_{ijk}=\tilde V_{ij}\tilde V_{jk}\tilde V_{ki}$ is equal to $\pm 1$
for all $i,j,k$ (since it equals $+1$ if projected to $\Spin(32)/\Z_2$),
obeys $W_{ijk}W_{jkl}W_{kli}W_{lij}=1$ in $U_i\cap U_j\cap U_k\cap U_l$,
and changes by a coboundary if the liftings $\tilde V_{ij}$ are changed.
Hence $W$ defines an element of $H^2(M,\Z_2)$, and this element
is $\tilde w_2(V)$.}

\nref\gimonpol{E. G. Gimon and J. Polchinski, ``Consistency Conditions
For Orientifolds And $D$-Manifolds,'' Phys. Rev. {\bf D54} (1996) 1667,
hep-th/9601038.}
\nref\gimonjohn{E. G. Gimon and C. V. Johnson, ``K3 Orientifolds,''
Nucl. Phys. {\bf B477} (1996) 715, hep-th/9604129.}
\nref\many{M. Berkooz, R. G. Leigh, J. Polchinski, J. H. Schwarz, N. Seiberg,
and E. Witten, ``Anomalies, Dualities, and Topology Of $D=6$ $N=1$
Superstring Vacua,'' Nucl. Phys. {\bf B475} (1996) 115.}
\nref\aspinwall{P. S. Aspinwall, 
``Pointlike Instantons And The $\Spin(32)/\Z_2$
Heterotic String,'' Nucl. Phys. {\bf B496} (1997) 149, hep-th/9611137;
P. S. Aspinwall and D. R. Morrison, ``Point-like Instantons On K3
Orbifolds,'' Nucl. Phys. {\bf B503} (1997) 533, hep-th/9705104.}
One reason for study of compactification without vector structure
is that many of the simplest orientifold constructions, such as K3 models 
constructed in
\refs{\gimonpol,\gimonjohn} can be interpreted as compactifications
without vector structure.
Also,  dualities can relate more familiar string
compactifications to compactifications without vector structure.
This has been seen \refs{\many,\aspinwall}
in studies of compactification on K3 surfaces.  
The intent of the present paper is
primarily to study compactification without vector structure in
the more elementary case of compactification on $\T^2$.
We will also, in less detail, study certain related and analogous
models in other dimensions.

In fact, the model we will focus on
 has been studied, from a rather
different vantage point (not stressing the topology)
in \ref\sagnotti{M. Bianchi, G. Pradisi,
and A. Sagnotti, ``Toroidal Compactification And Symmetry Breaking 
In Open String Theories,'' Nucl. Phys. {\bf B376} (1992) 365.}.
A number of important features were pointed out in that work, 
including the continuous
interpolation from $Sp(8)$ to $SO(16)$ and the role of a discrete theta
angle.  The model has also been investigated recently in 
\ref\lerche{W. Lerche, C. Schweigert, R. Minasian, and S. Theisen,
``A Note On The Geometry Of CHL Heterotic Strings,'' hep-th/9711104.}
and in \ref\bianchi{M. Bianchi, ``A Note On Toroidal Compactifications Of
The Type I Superstring And Other Superstring Vacuum Configurations With
16 Supercharges,'' hep-th/9711201.},
which appeared while the present paper was being written and have some
overlap with it.

Since the $\Spin(32)/\Z_2$ heterotic string and the Type I superstring
are equivalent already in ten dimensions, any of its
compactifications can be studied in either formalism.  This is so
whether there is vector structure or not.  In addition, for
the conventional case with vector structure, there are
the following important dualities that arise upon compactification
on $\T^2$:

\nref\dlp{J. Dai, R. G. Leigh, and J. Polchinski,
``New Connections Between String Theories,'' Mod. Phys. Lett.
{\bf A4} (1989) 2073.}
\nref\horava{P. Horava, ``Strings On World Sheet Orbifolds,'' Nucl. Phys.
{\bf B327} (1989) 461, ``Background Duality Of Open String Models,''
Phys. Lett. {\bf B231} (1989) 251.}
(1) Starting with the Type I superstring on $\T^2$, one can make
a $T$-duality transformation to a Type IIB orientifold on $\T^2/\Z_2$
with 16 sevenbrane pairs and 4 orientifold planes
\refs{\dlp,\horava}.

(2) In the strong coupling limit of the heterotic string, one gets
a description via $F$-theory on K3
\ref\vafa{C. Vafa, ``Evidence For $F$ Theory,'' Nucl. Phys. {\bf B469}
(1996) 403, hep-th/9602022.}, which can also be obtained as a strong coupling
limit \ref\sen{A. Sen, ``$F$ Theory And Orientifolds,'' Nucl. Phys. {\bf B475}
(1996) 562.} of the orientifold  as seen by
a threebrane probe \ref\probe{T. Banks, M. R. Douglas, and N. Seiberg,
``Probing $F$ Theory With Branes,'' Phys. Lett. {\bf B387} (1996) 278,
hep-th/9605199.}.

(3) Finally, via a heterotic string $T$-duality transformation,
one can map to a standard $\T^2$ compactification of the
$E_8\times E_8$ heterotic string \nref\narain{K. S. Narain, ``New Heterotic
String Theories In Uncompactified Dimensions $<10$,'' Phys. Lett.
{\bf B169} (1986) 41; K. S. Narain, M. H. Sarmadi, and E. Witten, ``A Note
On Toroidal Compactification,'' Nucl. Phys. {\bf B279} (1987) 369.}\nref\gins{
P. Ginsparg, ``Comment On Toroidal Compactification Of Heterotic
Superstrings,'' Phys. Rev. {\bf D35} (1987) 648.} \refs{\narain,\gins}.

Each of these dualities plays an important role in understanding
conventional $\T^2$ compactification of the $SO(32)$ heterotic
string.  As we will see, they all have analogs for 
compactification without vector structure: 

($1'$) Type I compactification on $\T^2$ without vector structure is
$T$-dual to a Type IIB orientifold on $\T^2/\Z_2$.  However,
in contrast to the usual case, there are only 8 sevenbrane pairs, 
and  of the  orientifold planes,  three have sevenbrane
charge $-8$ and one has charge $+8$.  

($2'$) $\T^2$ compactification without vector structure is equivalent
to a slightly exotic version of $F$-theory compactification on K3 in 
which the K3 surface is required to have a $D_8$ singularity that
does not produce gauge symmetry.  (This result is closely related
to a recent result \ref\ll{K. Landsteiner and E. Lopez,
``New Curves From Branes,'' hep-th/9708118.} 
about Type IIA orientifold sixplanes,
as we explain in section 4.3.)

($3'$) Finally, by a heterotic string $T$-duality transformation,
$\T^2$  compactification without vector structure can be mapped to
an $E_8\times E_8$ compactification on $\T^2$ in which the two
$E_8$'s are swapped in going around one circle in $\T^2$.  This
model was studied in \ref\chaud{S. Chaudhuri  and J. Polchinski,
``Moduli Space Of CHL Strings,'' Phys. Rev. {\bf D52} (1995) 7168,
hep-th/9506048.}, and argued there,
in the concluding paragraph of section 2.1, to be equivalent
to a $\Spin(32)/\Z_2$ model, which is in fact the one without vector
structure.  This relationship has recently been discussed in more detail in 
\lerche.

Dualities (3) and $(3')$ in the above lists have the further implication
that $\T^2$ compactification of the Type I superstring, with or without
vector structure, has an $M$-theory
description.  Indeed, the $E_8\times E_8$ heterotic string in ten dimensions
is equivalent
to $M$-theory on $\R^{10}\times\S^1/\Z_2$, so all its compactifications
likewise have $M$-theory descriptions.  The $M$-theory interpretation
of $\T^2$ compactification without vector structure will not be much
explored in the present paper, but will make a brief appearance in
section 5.

\nref\newj{
J. Polchinski, ``TASI Lectures on D-Branes,'' in {\it Fields, Strings,
Duality, TASI 96,} ed. C. Efthimiou and B. Greene (World
Scientific, Singapore, 1997), hep-th/9611050;
J. Polchinski, S. Chaudhuri, and C. V. Johnson, ``Notes on D-Branes,''
hep-th/9602052.}
By studying toroidal compactification without vector structure,
we will get a new insight about many familiar features of orientifolds,
such as the existence of different kinds of orientifold plane with
orthogonal or symplectic gauge symmetry and the
interplay between orthogonal and symplectic gauge symmetry on different
kinds of probe, as reviewed in \newj.

We begin in section 2 by studying $\T^2$ compactification without
vector structure at the level of classical gauge theory.  In the process
we uncover many interesting facts that, in  the stringy dualities
($1'$), ($2'$), and ($3'$), will appear  in different ways.
These include the fact that $\T^2$ compactification without vector
structure gives a gauge group whose rank is smaller by 8 than one
gets in the usual case with vector structure, and that it gives
simply laced gauge groups at level two and non-simply laced groups
at level one.  In sections 3-5, we study the three stringy dualities
listed above.

In section 3, we also briefly examine $\T^4$ and $\T^6$
compactification without vector structure.  

In section 6 we study a related question suggested by this
investigation.  There are actually {\it three} supersymmetric
Type IIB orientifolds
on $\T^2/\Z_2$.  The case in which all four orientifold planes
have sevenbrane charge $-8$ is $T$-dual to Type I on $\T^2$ with
vector structure \refs{\dlp,\horava}, and 
the present paper is largely devoted to showing that 
the case of three such planes
of charge $-8$ and one of charge $+8$ is $T$-dual
to the Type I string without vector structure.  This leaves a third
case, in which two orientifolds have charge $-8$, 
two have charge $+8$, and the number of sevenbranes is therefore
zero.  (This is the last model of its type, since 
if the net sevenbrane charge of the orientifolds is positive, one could
restore neutrality only by adding anti-sevenbranes and violating 
supersymmetry.)
In section 6, we study this model, show that it arises
by dimensional reduction from a {\it nine}-dimensional model with orientifold
planes of opposite type, and find the analogs of ($1'$) and ($2'$).

Finally, in an appendix (which can be read independently of the rest of
the paper) we resolve a longstanding puzzle concerning
the computation of $\Tr\, (-1)^F$ in four-dimensional supersymmetric
gauge theories with orthogonal gauge groups.

\newsec{Gauge Theory Analysis}

\subsec{Maximal Unbroken Symmetries}

We will begin in this section
by analyzing classical flat connections on $\T^2$
without vector structure.\foot{For background on some parts of this
discussion, see \ref\schwei{C. Schweigert,
``On Moduli Spaces of Flat Connections With Non-simply 
Connected Structure Group,'' hep-th/9611092.}.}
In doing so, we will initially take the gauge group to be $SO(32)/\Z_2$
(with the $\Z_2$ generated by the element $-1\in SO(32)$), and only
at the end take the double cover from $SO(32)/\Z_2$ to $\Spin(32)/\Z_2$.
Furthermore, to make the exposition  somewhat clearer, we first consider
$SO(4n)/\Z_2$ for general $n$, and only at the end specialize to $n=8$.

Given a flat connection on $\T^2$ with values in any gauge group,
one has holonomies $U,V$ around the two factors in $\T^2=\S^1\times\S^1$.
In our case, absence of vector structure means that $U$ and $V$ commute
as elements of $SO(4n)/\Z_2$, but anticommute in $SO(4n)$:
\eqn\uggu{UV=-VU.}
Let us first give certain important examples of such anticommuting
pairs.

$SO(4n)$ has a maximal subgroup $(Sp(1)\times Sp(n))/\Z_2$,
where the $\Z_2$ is generated by the product of $-1\in Sp(1)$
and $-1\in Sp(n)$.  Under this decomposition, the vector
representation of $SO(4n)$ decomposes as ${\bf 4n}={\bf 2}\otimes {\bf 2n}$,
where ${\bf 2}$ and ${\bf 2n}$ are, respectively, the fundamental
representations of $Sp(1)$ and $Sp(n)$.
Furthermore, we have $Sp(1)\cong SU(2)$. Up to conjugation, there is
a unique choice of $SU(2)$ matrices $u,v$ with $uv=-vu$:
\foot{ The relation $vuv^{-1}=-u$
shows that $u$ must be traceless and hence when diagonalized takes
the form given in the text.  Then, the fact that $v$ anticommutes with
$u$ means that its diagonal matrix elements vanish in this basis;
so up to conjugation by a diagonal matrix, $v$  takes the claimed form.}
\eqn\norgo{\eqalign{u & = \left(\matrix{i & 0 \cr 0 & -i \cr}\right)\cr
 v & = \left(\matrix{0 & 1 \cr -1 & 0 \cr}\right).\cr}}
By taking $U$ and $V$ to be the $SO(4n)$ matrices
$U=u\times 1$, $V=v\times 1$ (where $u\times 1$, for instance, is
the product of $u\in SU(2)$ 
with the identity in $Sp(n)$), 
we get an example of an $SO(4n)$  flat connection
without vector structure.  

The importance of this example is that it gives a maximal unbroken
subgroup of $SO(4n)$; in fact, the given $U$ and $V$ obviously commute
with $Sp(n)$.  The unbroken symmetry group is actually somewhat
bigger than this, because for an element $g\in SO(4n)$ to project
to an $SO(4n)/\Z_2$ transformation that commutes with the projection of
$U$ and $V$, it is enough for $g$ to commute {\it or anticommute}
with $U$ and $V$.  This gives an extra $\Z_2\times \Z_2$ factor in the
unbroken symmetry group, since not only does $1\times x$ (for any
$x\in Sp(n)$) commute
with $U=u\times 1$ and $V=v\times 1$, but $u\times x$, $v\times x$,
and $uv\times x$ commute or anticommute with them.  We will somewhat
imprecisely refer to the subgroup of $SO(4n)$ that commutes with the Wilson
lines as the unbroken symmetry group, and thus will suppress from the
terminology this $\Z_2\times \Z_2$ which is in fact always present.  (The
$\Z_2\times \Z_2$ will reappear in section 3 as the basis for one explanation
of why the dual torus has half the usual size.)

Presently, we will classify the possible unbroken subgroups in the above
sense, and it will become clear
that $Sp(n)$ is maximal in the sense that no $SO(4n)$
flat connection without vector structure has a symmetry group
that contains $Sp(n)$ as a proper subgroup.
However, $Sp(n)$ is not the unique maximal unbroken subgroup.
Another possibility is $O(2n)$.

$O(2n)$ can arise in the following way.  $SO(4n)$ has a maximal
subgroup $(O(2)\times O(2n))/\Z_2$ (with $\Z_2$ generated by the
product of $-1\in O(2)$ and $-1\in O(2n)$) under which the vector
representation of $SO(4n)$ decomposes as ${\bf 4n}=
{\bf 2}\otimes {\bf 2n}$, with
the two factors now the vector representations of $O(2)$ and $O(2n)$.  
The relation $uv=-vu$
can be obeyed by the  $O(2)$ matrices
\eqn\nokko{\eqalign{u & = \left(\matrix{1& 0 \cr 0 & -1 \cr}\right)\cr
 v & = \left(\matrix{0 & 1 \cr -1 & 0 \cr}\right).\cr}}
By setting $U=u\times 1$, $V=v\times 1$, we get a flat $SO(4n)$ bundle
without vector structure with unbroken gauge group $O(2n)$.
$O(2n)$ cannot be embedded in $Sp(n)$, and is a new example of
a maximal symmetry group.

The $Sp(n)$ and $O(2n)$ constructions differ from each other in the following
way.  Up to a gauge transformation, there is only one flat $SO(4n)$ bundle
without vector structure with 
unbroken $Sp(n)$.  This is so because the structure group of such
a bundle would reduce to $SU(2)$ (the commutant of $Sp(n)$), but
 anticommuting
matrices $u,v\in SU(2)$ are unique up to conjugation (as was
proved in the footnote accompanying \norgo).  However, in $O(2)$
it is not true that the choice of group elements $u,v$ with $uv=-vu$
is unique up to gauge transformation.   With the specific choice in \nokko,
we have $\det u=-1$, $\det v=+1$.  These conditions are not invariant
under replacing $(u,v)$ by $(v,u)$ or $(uv,v)$ (operations that do preserve
$uv=-vu$), so we get at least
three gauge-inequivalent flat bundles with unbroken $O(2n)$.
Up to conjugation there are precisely
these three possibilities; this may be proved as follows.
Consider any flat $O(2)$ connection on $\T^2$ without vector structure.
Let $\Gamma$ be a two-dimensional lattice with $\T^2=\R^2/\Gamma$.
Define a homomorphism  $\phi:
\Gamma\to\Z_2$ (with $\Z_2$
regarded as the multiplicative group $\{\pm 1\}$)
by mapping each $\gamma\in \Gamma$ to the determinant of the holonomy
of the given $O(2)$ connection around the one-cycle in $\T^2$ that
corresponds to $\gamma$.  
The homomorphism $\phi$ must be non-trivial; if it were trivial, the flat
$O(2)$ connection would actually be an $SO(2)$ connection, but as $SO(2)$
is abelian this would make the relation $uv=-vu$ impossible.
There are three non-zero possibilities for $\phi$; 
 the three possibilities are that the restrictions of $\phi$ to
 a basis of the lattice
can be $(-1,1)$, $(1,-1)$, or $(-1,-1)$.  An
$O(2)$ flat connection without vector structure is uniquely
determined up to gauge transformation by its associated $\phi$.  
For example, in the $(-1,1)$ case, 
 the holonomy $u$ around the first circle
  has determinant $-1$, and so $u$ is conjugate to the matrix given
in \nokko.  With this choice of $u$, the fact that the holonomy
$v$ around the second circle is of determinant
1 and anticommutes with $u$ leads (up to conjugation by $u$) to
the choice in \nokko.   

The mapping class group $SL(2,\Z)$ of $\T^2$
acts on the moduli space of flat connections without vector structure.
The uniqueness of the point
with unbroken $Sp(n)$ implies that it is $SL(2,\Z)$-invariant.
But the three points with unbroken $O(2n)$ are permuted by $SL(2,\Z)$
like the $\phi$'s, that is, like
the three real line bundles of order two on $\T^2$
(or equivalently, like the half-lattice points of $\Gamma$).

Maximal symmetry groups other than those that we have seen so far
can be obtained by starting
with $SO(4k)\times  SO(4(n-k))\subset SO(4n)$, and making one of the
two constructions that we have so far seen
in $SO(4k)$ and the other in $SO(4(n-k))$.  In this way we get
flat bundles with unbroken symmetry $Sp(k)\times O(2(n-k))$ for
any $k$.  As we will prove below, this is the complete list of 
possible maximal unbroken symmetry groups.  However, if we define
a locally maximal unbroken symmetry to be the symmetry of a flat
connection that has the property that under any small perturbation the
gauge symmetry would be reduced, then there are
other locally maximal examples.  They can be built
starting with $\prod_{i=1}^4SO(4k_i)\subset SO(4n)$, with
$\sum_{i=1}^4 k_i=n$, and making the $Sp$ construction in the first
factor and the three $SO$ constructions in the other three factors.
It will soon be clear that all of these statements have an intuitive
explanation in terms of sevenbranes. 

\subsec{Systematic Classification}

It follows from standard theorems about flat connections on Riemann
surfaces that the moduli space of flat connections on $\T^2$ without
vector structure is connected,\foot{The equivalence between flat
unitary connections and stable holomorphic bundles on a Riemann surface
means that a flat
connection $A$ can be reconstructed up to gauge transformation
from its $\bar\partial$ operator $\bar\partial_A$.  Given two
flat connections $A$ and $A'$ on the same bundle, one can continuously
interpolate between the $\bar\partial$ operators by looking at the
family $t\bar\partial_A+(1-t)\bar\partial_{A'}$, with $t$ varying from
0 to 1; this interpolation between the $\bar\partial$ operators induces
an interpolation between the flat connections.  It is possible to make
this interpolation because in complex dimension one there is no
integrability condition for $\bar\partial$ operators.  The absence
of obstructions to the deformation of a $\bar\partial$ operator in complex
dimension one also implies that the moduli spaces of flat unitary connections
are irreducible (they do not have different components meeting on a 
submanifold).}
so it must be possible to continuously interpolate between the
examples that we have given of flat connections without 
vector structure.  We will now describe this
interpolation explicitly, and in the process will see that
the possible maximal symmetry groups of a flat connection
without vector structure are those listed in the last
paragraph.

Suppose that we start at a flat connection (on an $SO(4n)/\Z_2$ bundle
without vector structure) with unbroken $Sp(n)$, and consider a small
perturbation of the flat connection. 
Any small perturbation can be made by introducing Wilson lines
of the unbroken subgroup, in this case $Sp(n)$. So
let $u',v'$ be any {\it commuting} elements of $Sp(n)$, and
let $U=u\times u'$, $V=v\times v'$, where $u,v$ are anticommuting elements
of $SU(2)$.  This gives a family of flat connections without vector
structure.

Since $u'$ and $ v'$ commute, they can be simultaneously conjugated
to a maximal torus of $Sp(n)$.  Such a torus is contained in a subgroup
$Sp(1)^n$ of $Sp(n)$.  As elements of $Sp(1)^n$, we can write $u'=\prod_{i=1}^n
 u_i$,
$v'=\prod_{i=1}^nv_i$, where $u_i$ and $v_i$ are the $Sp(1)$ or $SU(2)$
elements
\eqn\uggu{\eqalign{u_i&=\left(\matrix{e^{i\theta_i}& 0 \cr
           0 & e^{-i\theta_i}\cr}\right)\cr
           v_i&=\left(\matrix{e^{i\psi_i}& 0 \cr
           0 & e^{-i\psi_i}\cr}\right).\cr}}
It will sometimes be convenient to combine $\theta_i$ and $\psi_i$
in the pair $a_i=(\theta_i,\psi_i)$.
The Weyl group of $Sp(n)$ acts by permutations of the $Sp(1)$ factors
and by Weyl transformations of the individual factors.  The permutations
act by permutations of the $i$ index of $a_i$,
and the $Sp(1)^n$ Weyl transformations act by $a_i\to \pm a_i$, 
with independent choices of sign for each $i$.
Without this sign ambiguity, we would interpret each $a_i$
as a point on 
a dual torus $\tilde\T^2$ (which can be interpreted as the moduli space of 
$U(1)$
flat connections on $\T^2$).  Modulo the sign ambiguity, each $a_i$
 determines a point on an orientifold $\tilde\T^2/\Z_2$.
The complete set $a_1,a_2,\dots, a_n$ modulo permutations
 is a collection of  $n$ unordered points on $\tilde \T^2/\Z_2$.
This is in fact the moduli space of flat $Sp(n)$ connections on $\T^2$.

However, the map from the $Sp(n)$ moduli space to the $SO(4n)/\Z_2$
space, given by $u',v'\to U=u\times u', \,V=v\times v'$, is not
one-to-one.  This map is surjective and is
locally an isomorphism at a generic point in the
moduli spaces,\foot{Upon picking a complex structure on $\T^2$
and identifying the moduli spaces of flat connections with moduli
spaces of stable bundles, the map between $Sp(n)$ and $SO(4n)/\Z_2$
moduli spaces is holomorphic.  Since this map is also an isomorphism near
$u'=v'=1$, and the target space is irreducible according to the previous
footnote, the map is surjective and in fact the $Sp(n)$ moduli space
is a finite cover of the $SO(4n)/\Z_2$ moduli space.  The covering
group is determined by the discrete identifications that are explained
momentarily.} but is a many-to-one map because there
are discrete identifications in $SO(4n)/\Z_2$ that would not be present
in $Sp(n)$.  

These discrete identifications are independent shifts
of $\theta_i$ and $\psi_j$ by $\pi$, generating in all a group $\Theta$ with
$2^{2n}$ elements.  To see these discrete identifications, note
that $SO(4n)$ contains a subgroup $(SO(4))^n$.  We identify $SO(4)$
with $(SU(2)\times SU(2))/\Z_2$, and refer to the factors as the ``first''
and ``second'' $SU(2)$'s.
We align the $(SO(4))^n$ subgroup with the $Sp(1)\times Sp(n)$ that was
used earlier in such a way that $Sp(1)$ is a diagonal subgroup of the
product of the $n$ ``first'' $SU(2)$'s, and $Sp(1)^n\subset
Sp(n)$ is the product of the $n$ ``second'' $SU(2)$'s.  In this setup, the 
group 
element $U=u\times u'$, with $u'=\prod_iu_i$, is the  element $U=\prod_i
(u\times u_i)$ of 
$(SO(4))^n$, and likewise we can write $V=\prod_i(v\times v_i)$. 
Now, let $W_k$ be the element of $SO(4)^n$ whose $k^{th}$
factor is $v\times 1$, while the other factors are 1.  Then conjugation
by $W_k$ shifts $\theta_i$ by  $\pi\delta_{ik}$ while leaving $\psi_j$        
invariant.  Similarly, if $X_k$ is the element of $SO(4)^n$ whose
$k^{th}$ factor is $u\times 1$ while the others are   1, then conjugation
by $X_k$ shifts $\psi_j$ by $\pi\delta_{jk}$ and leaves $\theta_i$
unchanged.  These transformations thus generate the full group $\Theta$
of independent $\pi$ shifts of all components of the $a_i$.
(A more intuitive explanation of the origin of the group $\Theta$
is that it comes by combining the Weyl groups of the various $Sp(n)$
and $O(2n)$ maximal unbroken symmetry groups.)

The fundamental domain of $\theta_i$, $\psi_j$, subject to these symmetries
can be taken to be
\eqn\ikko{\eqalign{ -\half \pi\leq & \theta_i \leq \half\pi\cr
                    -\half\pi\leq & \psi_j \leq\half\pi.\cr}}         
Since the endpoints at $\pm \pi/2$ are identified, the $\theta_i$
and $\psi_j$ are still angular variables, but with half the usual
period.  Hence, each $a_i$ defines a point in a torus, which we will
call $\tilde \T^2$,
but (despite our unchanged notation) this is
{\it not} the usual dual torus; it is twice as small in each direction,
and its area is one-fourth the area of the usual dual torus.  
We still have to divide by the Weyl group of $Sp(n)$, but this can be
done just as in the case of ordinary $Sp(n)$ flat connections that
was treated earlier.  The sign changes $a_i\to -a_i$ 
(with independent signs for each $i$) mean that each  $a_i$ 
defines a point on an orientifold $\tilde \T^2/\Z_2$.
Including the $i$ index and dividing by permutations, the moduli space
of $SO(4n)/\Z_2$ connections without vector structure is the configuration
space of $n$ unordered 
points on $\tilde \T^2/\Z_2$.  Those points will
 be interpreted as positions of sevenbranes, roughly as
in the more familiar case \refs{\dlp,\horava} with vector structure.

Finally, let us identify the unbroken gauge symmetry group for
each set of sevenbrane positions.  Generically, the unbroken symmetry
group is a subgroup of $Sp(n)$.  When it is such a subgroup, it is
easy to determine which one.  For completely generic $a_i$,
the unbroken subgroup of the gauge group is a maximal torus $U(1)^n$
of $Sp(n)$.  If $k$ of the $a_i$ coincide at a generic point,
the associated $U(1)^k$ is enhanced to a $U(k)$ subgoup of $Sp(n)$, just
as for standard orientifolds.  If $k$ of
the $a_i$ coincide at $a_i=0$, the associated $U(1)^k$ is enhanced
to an $Sp(k)$ subgroup.

The interesting phenomenon that remains is that for certain special
(and non-zero) values of the $a_i$, there is an unbroken symmetry
group that is {\it not} a subgroup of $Sp(n)$.  
This happens, in fact, precisely when some of the $a_i$ are at orientifold
fixed points in $\tilde \T^2/\Z_2$ that are {\it not} at
$a_i=0$.  To analyze this, we can proceed as follows.  The part of
the Lie algebra of $SO(4n)$ that is not in the Lie algebra of $Sp(1)
\times Sp(n)$ transforms in the representation $R={\bf 3}\otimes {\bf A}$,
where ${\bf 3}$ is the adjoint representation of $SU(2)$ and ${\bf A}$
is the traceless second rank antisymmetric tensor of $Sp(n)$.  
Enhanced gauge symmetry not contained in $Sp(n)$ will arise if and only
if there are non-zero vectors in $R$ that are invariant
under the holonomies $U$ and $V$.  Let us determine the condition for this.

The ${\bf 3}$ of $SU(2)$ can be identified as the vector representation of
$SO(3)$, and if we think about it this way, then $u$ and $v$ (which
commute in $SO(3)=SU(2)/\Z_2$) can be identified with diagonal matrices
\eqn\huggob{\eqalign{u & =\left(\matrix{-1 & 0 & 0 \cr 0 & 1 & 0 \cr
   0 & 0 & -1\cr}\right)   \cr
   v& =\left(\matrix{1 & 0 & 0 \cr 0 & -1 & 0 \cr
   0 & 0 & -1\cr}\right)   .\cr}}
Since these matrices are diagonal in this basis, the subspace of
$R$ that is invariant under $U$ and $V$ is a direct sum of joint
eigenspaces of $u$ and $v$.  For instance, we can look for vectors
$\psi$ with $u\psi=-\psi$, $v\psi=\psi$.

For such a $\psi$ to be invariant under $U$ and $V$,
it must obey also $u'\psi=-\psi$, $v'\psi=\psi$.
Let us analyze these conditions.  Let ${\bf 2}_i$ be the two-dimensional
representation of the $i^{th}$ copy of $Sp(1)$ in $(Sp(1))^n\subset
Sp(n)$.  Then the $Sp(n)$ representation ${\bf A}$ is the sum
$\oplus_{1\leq i<j\leq n}{\bf 2}_i\otimes {\bf 2}_j$ plus 
``diagonal'' terms on which $Sp(1)^n$ acts trivially; the diagonal
terms cannot contribute vectors with $u'\psi=-\psi$.
The action of $U$ and $V$ is block-diagonal in the decomposition of ${\bf A}$
in terms of $\oplus_{1\leq i<j\leq n}{\bf 2}_i\otimes {\bf 2}_j$,
so  $\psi$ can be
taken to be a sum of vectors in ${\bf 2}_i\otimes {\bf 2}_j$.
The existence of a non-zero vector $\psi\in {\bf 2}_i\otimes {\bf 2}_j$ 
with $u'\psi=-\psi$, $v'\psi=\psi$
implies that (modulo sign changes of the $\theta$'s 
generated by Weyl transformations)
\eqn\indigo{ \eqalign{\theta_i=\theta_j &=\pi/2\cr
     \psi_i& = -\psi_j .                 }}
Such a configuration can actually be mapped to $a_i=a_j$,
which gives enhanced gauge symmetry in $Sp(n)$,
by a transformation in $\Theta$ (namely $\theta_j\to \theta_j+\pi$)
plus a Weyl transformation (changing the sign of $a_j$).
Hence, for generic $\psi_i$ and $\psi_j$, this mechanism gives
nothing essentially new.  But if we actually have $\psi_i=\psi_j=0$,
then in addition to \indigo, we have $a_i=a_j$, giving enhanced gauge
symmetry {\it inside} $Sp(n)$ as well as the enhanced gauge symmetry
that appears because \indigo\ is obeyed.  
It is impossible by a $\Theta$ plus Weyl
transformation to map everything into $Sp(n)$, so from this configuration
we get 
something essentially new.
         
The relevant case is thus
\eqn\diggo{a_i=a_j=(\pi/2,0).}
Similarly, symmetry enhancement in other eigenspaces of $u,v$ that cannot
be conjugated into $Sp(n)$ is associated
with some of the $a$'s being located at one of the other nonzero
orientifold fixed points, namely $(0,\pi/2)$ or $(\pi/2,\pi/2)$.

If precisely $k$ of the $a_i$ lie at $(\pi/2,0)$ (or one of the other
non-zero orientifold fixed points), then the unbroken symmetry
subgroup of $Sp(n)$ associated with the $a_i$
is $U(k)$.  In addition, one gets
out of each of the $k(k-1)/2$ relevant ${\bf 2}_i\otimes{\bf 2}_j$'s,
precisely 
two unbroken generators that are {\it not} in $Sp(n)$.                       
By (for example) analyzing the action of a maximal torus of $U(k)$, it
can be seen that these transform as the second rank antisymmetric
tensor of $U(k)$ plus its dual, so the Lie algebra of the
unbroken symmetry group is
actually that of $SO(2k)$.  With more care, one can see that the unbroken
symmetry group is $O(2k)$.    

If then we set $k=n$, we get, for each of the non-zero orientifold
fixed points,  a flat connection without vector structure with
unbroken $O(2n)$.  These of course were written down by hand at the
beginning of the present section.  More generally, if one distributes
all $a_i$ among the four orientifold fixed points in an arbitrary
fashion, one gets the locally maximal unbroken symmetry groups mentioned
at the end of the section 2.1.

\subsec{Comparison With Standard Orientifold}

Now we want to set $n=8$, so that $SO(4n)$ becomes the $SO(32)$
of the heterotic or Type I superstring, and compare to the standard
analysis of orientifolds.

The description of the moduli space that we
have found is quite similar to the one that is associated with
the standard orientifold with vector structure \refs{\dlp,\horava}.
The moduli space of $\T^2$ compactification without vector
structure can be described in terms of branes
on an orientifold $\tilde {\bf T}^2/\Z_2$, but
there are some important differences from the usual case:

(1) The orientifold is half as large in each direction as in the usual
case, and has therefore only one-fourth the usual area.

(2)  There are only eight sevenbranes on the orientifold
(or eight pairs of sevenbranes on the covering space $\tilde \T^2$),
which is just  half the number  on the standard orientifold.

(3)   The four orientifold planes, which are derived
from the fixed points of the $\Z_2$ action on $\tilde \T^2$, are of
two different kinds.   If $k$ sevenbranes meet the orientifold
plane at $a=0$, then $Sp(k)$
gauge symmetry is produced, while $k$ sevenbranes at one of the other
three orientifold planes produce $O(2k)$ gauge symmetry.

(4)  The rank of the gauge group is reduced by eight compared
to what one has in standard $\T^2$ compactification.
While standard $\T^2$ compactification gives only simply-laced
symmetry groups at level one, here we get the non-simply-laced
$Sp(k)$ groups at level one and simply-laced $SU(n)$ or $SO(2n)$
at level two.

\def\O{{\cal O}}
 A few additional words of explanation concerning these points should
suffice.  The statement that the orientifold is half the usual size
(one-fourth the usual area) was obtained  in eqn. \ikko.
The fact that there are only eight sevenbranes, which is clear
from the rank of the unbroken gauge group, is correlated with the
fact that there are orientifold planes of both kinds.  Unoriented
strings can have $Sp $ or $SO$ Chan-Paton factors (for a
review see section 1.3 of \ref\schwarz{J. H.
Schwarz, ``Superstring Theory,'' Physics Reports {\bf 89} (1982) 223.};
see also \ref\nm{N. Marcus and A. Sagnotti, ``Tree Level Constraints
On Gauge Groups For Type-I Superstrings,'' Phys. Lett. {\bf 119B} (1982) 97.} 
concerning the restriction to classical groups), 
and related to this, their $T$-duals
can have two kinds of orientifold plane, which produce respectively
$Sp$ and $SO$ gauge symmetry (for reviews see 
\newj).
 We will call them $\O^-$ and $\O^+$.
  They produce tadpoles of opposite
sign; $\O^+$
and $\O^-$ sevenplanes have sevenbrane charges  $-8$ and $+8$
respectively.  The  $\tilde \T^2/\Z_2$ orientifold that is $T$-dual
to standard $\T^2$ compactification of Type I
  has four orientifold
planes of type $\O^+$.  In this case,
the orientifolds carry a total charge of $-32$, so one requires
16 pairs of sevenbranes.  In the present case, the appearance of
$Sp$ symmetry at one sevenbrane and $SO$ at the other three shows
that we have three $\O^+$ planes and one $\O^-$ plane, with a net
sevenbrane charge of $-16$.  We therefore should expect eight pairs
of sevenbranes instead of 16, and the rank of the unbroken subgroup
of $SO(32)$ should be 8.  This agrees with what we have
found from classical gauge theory (since for $SO(4n)/\Z_2$ the
unbroken gauge group has rank $n$).

Finally, thinking in terms of the
$\Spin(32)/\Z_2$ heterotic string, 
and considering only the gauge symmetries that arise for generic radius
(other cases were studied in \chaud),
the $Sp(k)$ gauge symmetry
in $\T^2$ compactification without vector structure is at level
one, since 
the embedding $Sp(k)\subset Sp(n)
\subset Sp(1)\times Sp(n)\subset SO(4n)$ is a level one embedding of
$Sp(k) $ in $SO(4n)$.
But the $U(k)$ and $SO(2k)$ gauge symmetries
are at level two,
since their embeddings in $SO(4n)$ by
 $U(k)\subset SO(2k)\subset SO(2n)\subset
O(2)\times O(2n)\subset SO(4n)$ are at level two.

Throughout this section, we have considered $SO(32)/\Z_2$ flat connections
without vector structure, while in the heterotic or Type I superstring
one really wants $\Spin(32)/\Z_2$.  The difference only arises if one
considers particles in spin representations of $\Spin(32)/\Z_2$.  These
are not seen in classical gauge theory, but arise in the Type I
description from Dirichlet one-branes wrapped on a cycle in $\T^2$,
which become one-branes on $\tilde \T^2/\Z_2$.  In a more precise
description that takes account of these states, we must distinguish two
configurations of sevenbranes that differ by a motion of any one sevenbrane
around a cycle in the orientifold.  The same remark applies
for the more familiar \refs{\dlp,\horava} orientifold that is related
to compactification with vector structure.

In this section, we have analyzed only classical gauge theory and not
string theory.  This has sufficed to see what kind of orientifold
must be $T$-dual to the Type I superstring on $\T^2$ without
vector structure.  In the next section, we will actually analyze
the $T$-duality and show how the expected answer arises.

\newsec{$T$-Duality}

\subsec{The Closed String Sector}

Our goal in the present section is to apply  $T$-duality to Type I
compactification on $\T^2$ without vector structure.  The aim is
to explain and recover  via $T$-duality the features that were
obtained by an analysis of classical gauge theory in section 2 and
summarized in section 2.3.  We especially want to explain why
the dual torus is of half the usual size (one quarter the usual
area) and why there are three
orientifold planes of type ${\cal O}^+$ and one of type ${\cal O}^-$.

We will begin with the closed strings,
which are of course unoriented, and then move on to the open strings.
At first sight we face a quandary.  The closed strings do not appear
to ``know'' whether there is vector structure or not, since
gauge fields appear only in the open string sector.  So how can
application of $T$-duality to the closed strings produce a torus
of half the size in case there is no vector structure?

What saves the day is that there is in fact a subtle correlation 
between the presence or absence of vector structure and the couplings
in the closed string sector.  We recall that for Type IIB superstrings,
there is a two-form that comes from the NS sector, and another from the
Ramond sector.  In compactification on $\T^2$, the NS two-form gives
rise to a world-sheet theta angle whose effects can be seen in
string perturbation theory.  (The Ramond two-form gives a second
theta-like angle with nonperturbative effects.)  The orientifold
projection to Type I removes the NS two-form; the Ramond two-form survives.
As for the world-sheet theta angle, it is odd under reversal
of world-sheet orientation, so in Type I it is severely restricted;
it must be 0 or $\pi$.

At first sight, it seems therefore that there are four Type I models
on $\T^2$ that we could potentially consider.  One may have or not
have vector structure; and the theta angle may be 0 or $\pi$.
However \ref\sethisen{S. Sethi and A. Sen, ``The Mirror Transform
Of Type I Vacua In Six-Dimensions,'' Nucl. Phys. {\bf B499} (1997) 45,
hep-th/9703157.}, these choices are correlated;
the theta angle is $\pi$ if and only if there is no vector structure.
So to study the theory without vector structure, we must set $\theta=\pi$.
In this way, the closed string sector does ``know'' that there is
no vector structure.  

Now, let us recall the structure of the $T$-duality group of $\T^2$.
It is $SO(2,2;\Z)\cong SL(2,\Z)\times SL(2,\Z)$.  The first $SL(2,\Z)$ acts
on the complex structure of $\T^2$ and will play practically no role
in what follows.  The second acts in the customary way on 
\eqn\ubb{\tau=iA+{\theta\over 2\pi}}
where $A$ is the area of $\T^2$ and $\theta$ is the theta angle.

What $SL(2,\Z)$ transformation do we want to make?  Application of
any $SL(2,\Z)$ transformation at all will give a correct result.
But our goal in applying $T$-duality will be to determine the behavior
of the model in the limit of small $A$, by finding an $SL(2,\Z)$
transformation that will map us back to large $A$.
At $\theta=0$, $\tau\to -1/\tau$ reduces to $A\to 1/A$, and so maps
small $A$ to large $A$.  But at $\theta=\pi$, $\tau\to -1/\tau$ maps
small $A$ to small $A$ and is not helpful.

A transformation that is more helpful at $\theta=\pi$ is
\eqn\imub{\tau\to {\tau-1\over 2\tau-1}.}
A short calculation reveals that at $\theta=\pi$, this transformation
reduces to 
\eqn\ikko{A\to {1\over 4A}}
(with no change in $\theta$), so it has the desired effect of mapping
small area back to large area.  We also see that, as anticipated via
classical gauge theory in section 2,  the area of the dual torus
is four times smaller than it is for the usual $T$-duality of $\T^2$.
Since the transformation \imub\ acts trivially on the complex structure
of the torus, the four-fold reduction in the area is achieved by
reducing all lengths by an extra factor of two compared to the standard
$T$-duality at $\theta=0$.

So we have obtained a ``stringy'' explanation of a result in section 2,
but why the particular $SL(2,\Z)$ transformation indicated in
\imub\ is the right one at $\theta=\pi$ begs for a more intuitive
explanation.  

\bigskip\noindent{\it Intuitive Explanation Of $R\to 1/(2R)$}

For this, we need to look at the $T$-duality more explicitly.
(See \ref\rab{A. Giveon, M. Porrati,
E. Rabinovici, ``Target Space Duality In String Theory,''
Phys. Rept. {\bf 244} (1994) 77.} 
for background and further detail. The rest
of this paper does not depend on the following intuitive explanation.)
We consider a rectangular $\T^2$ with radii $R_1$, $R_2$.  The usual
$T$-duality, at $\theta=0$, is $(R_1,R_2)\to (1/R_1,1/R_2)$.  It is convenient
to combine this with a $\pi/2$ rotation, to give
$(R_1,R_2)\to (1/R_2,1/R_1)=A^{-1}(R_1,R_2)$.  This makes it explicit
that this transformation inverts the area without changing the complex
structure.  

$T$-duality acts on a four-dimensional lattice
consisting of momenta $p_x, p_y$ along the two directions in $\T^2$
and corresponding windings $m_x,m_y$.  (We measure momenta and
winding in units such that the $p$'s and $m$'s are integral at
$\theta=0$.)
For our purposes, we can reduce to a two-dimensional lattice, as follows.
An $R\to 1/R$ transformation is $p_x\leftrightarrow m_y$
and $p_y\leftrightarrow -m_x$.  (Of course, $p_x$ is exchanged with
$m_y$ instead of $m_x$ since we combine the $R\to 1/R$ transformation
with a $\pi/2$ rotation; for the same reason $p_y$ maps to $-m_x$.)
This transformation, together with $\theta\to\theta+2\pi$, generates
the $SL(2,\Z)$ that acts on $A$ and $\theta$.  

At non-zero $\theta$, the $p$'s are shifted from their conventional
integral values.  In fact, $p_x$ is shifted
from an integer by $m_y(\theta/ 2 \pi)$, and $p_y$ by 
$-m_x(\theta/2\pi)$.  The transformation $\theta\to\theta+2\pi$
is hence $p_x\to p_x+m_y$, $p_y\to p_y-m_x$.  In particular,
the lattice generated by $p_x$ and $m_y$ (and likewise the lattice
generated by $p_y$ and $m_x$) is mapped to itself by  $\theta\to\theta+2\pi$
as well as by $R\to 1/R$, so it is mapped to itself by the $SL(2,\Z)$
that acts on $A$ and $\theta$.  We can focus just on the $p_x, m_y$ lattice.

Now let us see what happens at $\theta=\pi$.  The $p_x,m_y$
lattice at this value of $\theta$
is generated by the vectors $f=(1,0)$ and $e=(-1/2,1)$.
Clearly, such a lattice cannot have a symmetry that exchanges
$p_x$ with $m_y$, since such a symmetry would map $(-1/2,1)$, which
is in the lattice, to $(1,-1/2)$, which is not.  However, 
let $h=(0,2)$.  The lattice is generated by $f$, $h$, and $\half(f+h)$,
a set of vectors that is symmetric under $f\leftrightarrow h$, so
it has a symmetry $W$ that acts by 
$f\to h$, $h\to -f$.  (The minus sign ensures that the determinant
is 1, so that $W$ is in $SL(2,\Z)$.)  Thus $W$ maps $p_x$ to $m_y/2$.
  The factor of 2 is what we are looking for.
We start with a radius $R$ in the $x$ direction, so the $(1,0)$ state
has energy $1/R$.  If the $W$ transformation produces a dual circle
with radius $\tilde R$, then the $(0,2)$ state after duality has
energy $2\tilde R$.  So  $1/R=2\tilde R$,
that is $\tilde R=1/(2R)$.  So the natural symmetry at $\theta=\pi$
is $R\to 1/(2R)$, as we have already discovered in several different ways.

The transformation $W$, which maps $f\to h$ and $h\to -f$,
can be written in terms of the original basis $e,f$ as
 $f\to f+2e$, $e\to -f-e$.  This corresponds in the basis $(f,e)$
 to the $SL(2,\Z)$
matrix
\eqn\hucco{\left(\matrix{ 1 & -1\cr 2 & -1 }\right).}
On the upper half plane, this acts by $\tau\to (\tau-1)/(2\tau-1)$,
which is the transformation whose origin we wished to explain
intuitively.

\bigskip\noindent{\it Nature Of The Orientifold Planes}

So far, our discussion applies most naturally to {\it oriented}
closed strings at $\theta=\pi$.  Now, we must adapt the discussion
to {\it unoriented} closed strings, relevant to Type I.

For the same reasons as in \refs{\dlp,\horava}, 
projecting the closed strings on the original torus $\T^2$ onto states
invariant under reversal of orientation is equivalent in the $T$-dual
description to replacing the dual torus $\tilde\T^2$ by an orientifold
$\tilde\T^2/\Z_2$.  We recall that this is argued as follows.
If $p_L$ and $p_R$ are the left and right-moving momenta of a closed string,
then $R\to 1/R$ (or $R\to 1/(2R)$) acts by $T:p_R\to p_R,p_L\to -p_L$.
On the other hand, world-sheet
orientation reversal is $\Omega:p_L\to p_R,p_R\to
p_L$.  $\Omega$ transforms under $T$-duality to
 $T\Omega T^{-1}$, which acts by $p_L\to -p_L, p_R\to -p_R$.
But, since it acts in the same way on $p_L$ and $p_R$,
 this is a geometrical transformation.  In fact, it acts on the dual torus
by ``multiplication by $-1$''; projecting onto states 
invariant under it replaces $\tilde \T^2$ by $\tilde \T^2/\Z_2$.

There are four fixed points in the $\Z_2$ action on $\tilde \T^2$,
and each becomes an orientifold plane. Now, however, we wish to explain
in terms of closed string $T$-duality another important result of
section 2, which is that in the case without vector structure,
the orientifolds have a net sevenbrane charge of $-16$, compared to
$-32$ for the case with vector structure.  We will do this
by carefully comparing the worldsheet path integrals for the
\def\TT{\tilde{\bf T}}\def\RP{{\bf RP}}
case that the worldsheet topology is $\S^2$ or $\RP^2$.

Let $A$ be the area of the original $\T^2$, $A'$ the area
of the dual $\TT^2/\Z_2$ obtained by $R\to 1/R$ at $\theta=0$,
and $A''$ the area of the dual $\TT^2/\Z_2$ obtained by $R\to 1/(2R)$
at $\theta=\pi$.  (Thus, $A''=A'/4$, a fact that will be used presently.)
Let also $g$ be the string coupling constant in the original description,
$g'$ the string coupling constant in the $T$-dual at $\theta=0$, and
$g''$ the string coupling constant in the $T$-dual at $\theta=\pi$.

Before $T$-duality, the world-sheet path integral for a worldsheet
$\Sigma$ that is topologically $\S^2$ or $\RP^2$ is independent
of $\theta$, since a worldsheet of the given topology cannot
``wrap'' around $\T^2$.  We want to impose the condition that also after
$T$-duality, the path integral is the same whether we are at $\theta=0$
or $\theta=\pi$.  

First we consider the case that the worldsheet is $\S^2$.  
The dependence on the coupling and area is extremely simple.
Before $T$-duality, as there are no wrapping modes the worldsheet
path integral is simply proportional to $A$; and of course in genus
zero one has a factor of $g^{-2}$.  Likewise after $T$-duality
there are still no wrapping modes in genus zero, so the partition
function is proportional to 
$A'/(g')^2$ or $A''/(g'')^2$.  
All other factors are the same whether $\theta$ is 0 or $\pi$.  So
equality of the partition functions after $T$-duality gives
\eqn\jungle{{A'\over (g')^2}={A''\over (g'')^2}.}
Since $A''=A'/4$, we get
\eqn\hocco{g''={g'\over 2}.}
 
Now, let us compare the $\RP^2$ partition functions, after $T$-duality,
at $\theta=0$ and $\theta=\pi$.
For ${\bf RP}^2$, the partition function 
has only one power of inverse string coupling, instead of two.
Also, there is no factor of area in the partition function of
the orientifold, because the ``center of mass'' of the worldsheet
is always mapped to one of the orientifold planes.
The partition function for $\RP^2$ therefore contains a sum over
orientifold planes, but it is not simply proportional to the number
of such planes.  There are two kinds of orientifold planes, called
${\cal O}^+$ and ${\cal O}^-$ in section 2, which contribute
$\RP^2$ partition functions of equal and opposite signs.  If
therefore $n_+$ and $n_-$ are the number of ${\cal O}^+$ and ${\cal O}^-$
planes -- so that 
\eqn\hungery{n_++n_-=4}
regardless of the value of $\theta$ --
then the worldsheet path integral for $\RP^2$ is proportional
to $\Delta=n_+-n_-$.  Letting $\Delta'$, $\Delta''$ be the values of
$\Delta$ at $\theta=0$ and $\theta=\pi$, respectively, we have therefore
\eqn\ungery{{\Delta'\over g'}={\Delta''\over g''}.}
In view of \hocco, this gives $\Delta''=\half \Delta'$.  Since
the standard orientifold at $\theta=0$ has $n_+=4,$ $n_-=0$, and $\Delta'=4$,
it follows that the $\theta=\pi$ orientifold has $\Delta''=2$,
and hence $n_+=3$ and $n_-=1$.  The values of $n_+$ and $n_-$ are of
course in agreement with what we found in section 2 using classical
gauge theory.

\subsec{$T$-Duality For Open Strings}

So far, we have only considered $T$-duality in the closed string sector.
Our next task is to  analyze $T$-duality for open
strings.  First we consider ordinary open strings of the underlying
Type I model -- sometimes called strings that end on ninebranes.
Then,  we will analyze $T$-duality for Dirichlet onebrane
probes.  The case of fivebranes will be postponed to section 4, and will
serve as the stepping stone to an $F$-theory description.

We examine the ordinary open strings first in the vacuum with unbroken
$Sp(8)$.  The Chan-Paton factors at the end of an open string
transform in the ${\bf 32}$ of $SO(32)$, which is the
${\bf 2}\otimes {\bf 16}$ of $(Sp(1)\times Sp(8))/\Z_2\subset SO(32)$.
An actual open string has a charge in this representation at each
end, so the open string itself transforms as 
$({\bf 2}\otimes {\bf 16})\otimes(
{\bf 2}\otimes {\bf 16})\cong ({\bf 2}\otimes {\bf 2})\otimes ({\bf 16}\otimes
{\bf 16})$, which in particular (unlike the ${\bf 32}$ itself) is
a representation of the group $\Spin(32)/\Z_2$.

The ${\bf 2}\otimes {\bf 2}$ of $Sp(1)=SU(2)$ can be
decomposed in the usual way as ${\bf 1}\oplus {\bf 3}$.
Since this is a representation of $SO(3)=SU(2)/\Z_2$ (with no
need for the double covering to $SU(2)$), the $SU(2)$ Wilson lines $u,v$
commute in this representation, and so can be simultaneously diagonalized.
In fact, in the ${\bf 3}$ this diagonalization was already made in
eqn.
\huggob, while in the ${\bf 1}$, $u$ and $v$ of course have the common
eigenvalue $1$. Combining these results, the ${\bf 1}\oplus {\bf 3}$ has
the property that the eigenvalues of the pair $(u,v)$ run over
all possible pairs $(\pm 1, \pm 1)$, with each of the four combinations
of signs appearing precisely once.  An eigenvalue $-1$ for $u$ or $v$
shifts the possible momenta of a particle or string by half a unit.
The fact that all pairs $(\pm 1,\pm 1)$ appear with unit multiplicity
means that all half-integral shifts appear with the same multiplicity
as the unshifted momenta.  Hence, the momenta of the open strings
take values not in the usual momentum lattice $\Lambda$ but in
a rescaled lattice $\half\Lambda$.

After $T$-duality, the momentum lattice of the open strings is reinterpreted
as a lattice of windings for open strings that now obey Dirichlet
boundary conditions.  To obtain a lattice $\Lambda$ of windings, one
usually has open strings on a dual torus $\tilde \T^2=\R^2/\Lambda$.
In the present case, to get the desired rescaled winding
lattice $\half\Lambda$,
we must define $\tilde \T^2=\R^2/(\half\Lambda)$.  So we recover
the result that we have by now obtained in several other ways:
in compactification without vector structure, the dual torus has
one-half the usual size.

Since the open strings are unoriented, a winding $\lambda\in\half\Lambda$
must be identified with $-\lambda$.  (Also, a certain projection
must be made on the states with $\lambda=0$.)  In the usual way
\refs{\dlp,\horava}, this means that $\tilde\T^2$ must be replaced
by the orientifold $\tilde\T^2/\Z_2$.  Also, the vacuum with
unbroken $Sp(8)$
has precisely eight sevenbrane pairs at the ${\cal O}^-$  fixed point in 
$\tilde \T^2/\Z_2$.  We of course get the other maximal unbroken
symmetry, $O(16)$, if all sevenbranes are placed at an ${\cal O}^+$ fixed
point.

It is perhaps easier to understand this construction if we compare
to the usual case with vector structure.  In that case, one has
16 sevenbrane pairs.  If one wants the windings to take values in
$\half\Lambda$ and not $\Lambda$, one must divide the sevenbrane
pairs into four groups (of four each), and place one group at
each orientifold fixed point.  
In this case, the unbroken gauge group
is $SO(8)^4$, with one factor from each of the four groups of sevenbranes.
In particular, the unbroken group is not simple.
The other way to get a winding lattice $\half\Lambda$ is the
one followed in the theory without vector structure: take
the dual torus to be half as big.  In this case, one can get
the simple unbroken gauge group $Sp(8)$ or $O(16)$.

\bigskip\noindent{\it $T$-Duality For One-Brane Probes}

Our next target will be to analyze $T$-duality for Dirichlet onebrane
probes of this system.  

We begin with Type I on $\R^8 \times\T^2$ without vector structure.
Consider a Dirichlet onebrane whose world-volume is localized
at a (fluctuating) point on $\T^2$, and occupies a two-dimensional
subspace $\R^2$ of $\R^8$.  Such a onebrane behaves as a ``solitonic heterotic 
string,'' and is transformed to an elementary heterotic string under
heterotic/Type I duality.  Under $T$-duality to the $\tilde\T^2/\Z_2$
orientifold, it becomes a threebrane whose world-volume is
$\R^2\times\tilde\T^2/\Z_2$ (with a further subtlety that will appear).

The point that we wish to investigate is the following.
Consider the $\Spin(32)/\Z_2$ heterotic string on $\T^2$ without
vector structure.  For certain choices of Wilson line, as analyzed
in detail in section 2, one gets an unbroken $U(k)$ gauge symmetry.
As noted in section 2.3, this is a level two gauge symmetry,
meaning that on the worldvolume of an extended heterotic string
transverse to the $\T^2$, there will be a level two $U(k)$ current
algebra.

After heterotic/Type I duality, it follows that an extended Dirichlet
onebrane transverse to the $\T^2$ will have such a level two current
algebra.  $T$-duality to the orientifold implies that the same
will be true for a threebrane wrapped on $\tilde\T^2/\Z_2$.
Our goal is to understand what the ``level two'' property means
in terms of the threebrane.

In terms of the orientifold, $U(k)$ gauge symmetry appears
when $k$ sevenbranes coincide at a  point $P$ on the orientifold that
is not a fixed point.
\foot{If they coincide at certain orientifold fixed points, one gets
an enhanced $SO(2k)$ gauge symmetry, which is at level two in the
heterotic string description as explained in section 2.3, and is
at level two in the orientifold description for the  reason
given below.}
The threebrane wrapped on $\R^2\times \tilde \T^2$ intersects
the sevenbrane on the two-manifold $\R^2\times P$.  A standard
quantization of the $3\cdot 7$ open strings shows that there are
massless chiral fermions, in the fundamental representation of $U(k)$
because there are $k$ sevenbranes, propagating on $\R^2\times P$.  This
gives the desired $U(k)$ current algebra (which is with a suitable
choice of orientation left-moving since the massless fermions are chiral).  
The problem is to explain
why the current algebra is at level {\it two}.

The answer depends on an interesting detail about how the $T$-duality
to the orientifold acts on Dirichlet onebranes that are transverse
to the $\T^2$.  Such a one-brane is mapped to a threebrane that
wraps {\it twice} over $\tilde\T^2/\Z_2$, as we will explain
presently.  Because the wrapping
number is two, the local structure near $P$ is actually
that of {\it two}
threebranes intersecting transversely with $k$ sevenbranes,
as a result of which there are twice as many chiral fermion
zero modes, and one gets a level 
two current algebra.

The origin of the twofold wrapping is that, in the Type I description,
a Dirichlet onebrane localized on $\R^2\times P$, since it lives
at a single point $P\in \T^2$, does not ``see'' the $\Spin(32)/\Z_2$
Wilson lines.  The allowed winding numbers of the $1\cdot 1$ open strings
hence
are conventional integers, with no half-integral shifts resulting
from the lack of vector structure.  Hence, after $T$-duality to
the orientifold, one must get a threebrane wrapped on $\tilde\T^2/\Z_2$
with the property that the allowed momenta for $3\cdot 3$ open strings
are the same as they would be if the underlying Type I model had
vector structure.

But as we have by now extensively seen, the $\tilde\T^2/\Z_2$ that arises
when there is no vector structure is of one-half the usual size.
So the     quantum of momentum for a threebrane wrapped once
on $\tilde\T^2/\Z_2$ would be {\it twice} what we want.
The cure for this involves a mechanism that we have already used,
in reverse, in discussing the conventional open strings, and which
has entered many times in studies of branes and string dualities.
Ignoring for a moment the orientifold
projection, we start with {\it two} threebranes wrapping on $\tilde\T^2$.
There is a world-volume $U(2)$ gauge symmetry group.  We choose a $U(2)/\Z_2$
flat connection on $\tilde\T^2$ such that the monodromies $A,B$ obey
$AB=-BA$.  
This makes sense for the $3\cdot 3$ open strings 
since they transform in the adjoint
representation of $U(2)$.\foot{All other open strings actually present
in the theory also make sense, but there are some exotic details
in some cases, for example involving the fact that a string
transforming in a $U(2)$ representation that does not descend
to $U(2)/\Z_2$ can make sense if its center of mass is localized on
a submanifold of $\tilde\T^2$ on which the $U(2)/\Z_2$ bundle
is trivial.}  In the adjoint representation of $U(2)$, the matrices
$A$ and $B$ commute and can be simultaneously diagonalized, with the
result that the eigenvalues of $(A,B)$ are $(\pm 1,\pm 1)$, with
each possible pair occurring precisely once.  Since an eigenvalue
$-1$ shifts the corresponding momentum by half a unit, the result of this
is that the momentum lattice for the $3\cdot 3$ open strings is not the
conventional momentum lattice $\Gamma$ of $\tilde \T^2$ but is
$\half\Gamma$.  Because $\tilde\T^2$ is half the usual size of the dual
lattice, $\Gamma$ is twice the appropriate momentum lattice for the
$3\cdot 3$ open strings, and hence $\half\Gamma$ is the right lattice.

After the orientifold projection to $\tilde \T^2/\Z_2$, the windings
and momenta used to classify the states in the last two paragraphs
are only conserved modulo two.  Nonetheless, a correct $T$-duality
must match up the spectra in the weak coupling 
limit, making possible the analysis in the last
two paragraphs.  The remaining point
is that the orientifold projection on the threebranes must be such
that the Wilson lines $A,B$ exist.  This orientifold projection
breaks $U(2)$ to $SO(2)$ (since Type I $D$-strings have $SO$ and
not $Sp$ Chan-Paton factors) and must conjugate $A$ and $B$ to
$A^{-1}$ and $B^{-1}$ (since the $\Z_2$ acts as $-1$ on $\tilde \T^2$).
An orientifold projection of the $SO$ kind admits matrices
$A,B$ with the claimed properties.

\subsec{Compactification To Lower Dimensions}

In this subsection, we consider an issue that is of some interest,
though  outside the main theme of the present paper:  compactification
of the $\Spin(32)/\Z_2$ superstring on a torus $\T^n$ with $n>2$.

The possible $\Spin(32)/\Z_2$ bundles are classified by the
characteristic class $\tilde w_2\in H^2(\T^n,\Z_2)$.
$\tilde w_2$ can be regarded as a second rank antisymmetric tensor
of the group $SL(n,\Z_2)$.  Just like an ordinary 	real-valued
second rank antisymmetric tensor, it can  by the action
of $SL(n,\Z_2)$ be brought to a skew-diagonal form in which the only
non-zero matrix elements are the $1\cdot 2$, $3\cdot 4$, $\dots$, $(2k-1)
\cdot 2k$
components, for some $k$ with $2k\leq n$.  $k$ is thus the only invariant
of $\tilde w_2$, and up to the action of $SL(n,\Z)$ (the mapping
class group of $\T^n$, which by reduction modulo two induces
the action of $SL(n,\Z_2)$ on $\tilde w_2$), 
the topological type of the bundle is determined by $k$.

There is thus up to diffeomorphism only one type of non-trivial bundle
on $\T^2$ or $\T^3$,
 but on $\T^4$ a second case appears, with $k=2$, and on $\T^6$,
there is a third possibility with $k=3$.  The $k=2$ bundle on $\T^4$
is characterized by $\int_{\T^4}\tilde w_2^2\not= 0$ 
(where integration is understood as a pairing in mod two cohomology),
and for $k=3$ one   likewise has $\int_{\T^6}\tilde w_2^3\not= 0$.

Let us briefly examine the compactification on  $\T^4$ with $k=2$.  
A flat connection on a $\Spin(32)/\Z_2$
bundle $X$ with $\tilde w_2^2\not= 0$ is characterized, in 
a suitable coordinate system, by $SO(32)$-valued
Wilson lines $U,V, A,B$ (the holonomies around the four directions
in $\T^4$),
with $UV=-VU$, $AB=-BA$, while $U$ and $V$ commute with $A$ and $B$.
Based on our experience in section 2, we can readily construct
examples of such matrices in several ways:

\def\2{{\bf 2}}
\def\8{{\bf 8}}
(1) We embed $SU(2)\times SU(2)'\times SO(8)$
(here $SU(2)$ and $SU(2)'$ are simply two copies of $SU(2)$) in $SO(32)$, in
such 
a way that the vector of $SO(32)$ decomposes as $\2\otimes \2'\otimes \8$,
where $\2,$ $\2'$, and $\8$ are the standard representations of
the three factors.  Then we take $U=u\times 1\times 1$, $V=v\times 1\times 1$,
$A=1\times u\times 1$, $B=1\times v\times 1$, with $u$ and $v$ taken
from eqn. \norgo.  This gives a flat
connection on the bundle $X$ with unbroken symmetry
an $SO(8)$ group at level 4.

(2) We embed $SU(2)\times O(2)\times Sp(4)$ in $ SO(32)$
in such a way that the vector of $SO(32)$ decomposes as $\2\times \2'
\times \8$, where $\2$, $\2'$, and $\8$ are again the standard representations
of the three factors.  Then we take $U=u\times 1\times 1$,
$V=v\times 1\times 1$ with $u$ and $v$ as in  eqn. \norgo;
and we take $A=1\times u\times 1$, $B=1\times v\times 1$, with now
$u$ and $v$ taken from eqn. \nokko.  This gives a flat connection on
the bundle $X$ with the unbroken gauge symmetry 
being an $Sp(4)$ group at level 2.
This construction has $6=2\times 3$ variants, since the pair $U,V$ can be
exchanged with $A,B$, and also we recall from section 2.1 that
the construction in eqn. \nokko\ has three variants.

(3) Finally, we embed $O(2)\times O(2)'\times SO(8)$ in  $SO(32)$
(here $O(2)$ and $O(2)'$ 
are simply two copies of $O(2)$) in $SO(32)$, in such 
a way that the vector of $SO(32)$ decomposes as $\2\otimes \2'\otimes \8$,
where $\2,$ $\2'$, and $\8$ are the standard representations of
the three factors.  Then we take $U=u\times 1\times 1$, $V=v\times 1\times 1$,
$A=1\times u\times 1$, $B=1\times v\times 1$, with $u$ and $v$ taken
from eqn.
\nokko.  This gives a flat connection on the bundle $X$ with unbroken 
symmetry an $SO(8)$ group at level 4.  
There are $9=3\times 3$ variants of this construction, since the
construction in eqn. \nokko, which
we used twice, has three variants.

Adding the above, we see we have in all 10 constructions of flat
bundles with $SO(8)$ gauge symmetry, and 6 with $Sp(4)$ gauge
symmetry.  Along the lines of arguments in section 2, one can show that
these 16 bundles lie in one connected component of the moduli space of
flat connections on $X$.  Moreover, this component is parametrized
by the positions of four fivebrane pairs on the orientifold $\tilde \T^4/\Z_2$.
Enhanced gauge symmetry arises when all fivebranes are at one of
the 16 orientifold fixed points.  10 of the fixed points are of
${\cal O}^+$ type and produce $SO$ gauge groups, while 6 are of
${\cal O}^-$ type and produce $Sp$ gauge groups.  The $SO$ and $Sp$
orientifold points have respectively a fivebrane charge $-2$ or $+2$,
so the net fivebrane charge is $-8$, and four fivebrane pairs is
the right number to ensure overall neutrality.  Moreover, $T$-duality
of the $k=2$ model leads, along the lines of arguments that we have
seen for $k=1$,
to this orientifold with 10 ${\cal O}^+$ and 6 ${\cal O}^-$ fixed points.

Similarly, the $k=3$ model is related to a $\tilde\T^6/\Z_2$ orientifold
with $36$ ${\cal O}^+$ fixed points of threebrane charge $-1/2$,
$28$ ${\cal O}^-$ fixed points of threebrane charge $+1/2$, and
two threebrane pairs.  When all threebranes approach a ${\cal O}^+$ point,
one gets $SO(4)$ gauge symmetry at level 8, and when all threebranes
approach a ${\cal O}^-$ point, one gets $Sp(2)$ gauge theory at level 4.

The $k=2$ and $k=3$ models have gauge groups of rank reduced by 12 and
14, respectively, compared to standard toroidal compactification.
They appear to correspond (as suggested by Y. Oz and also in \bianchi) 
to the second and
third models in the CHL series, as enumerated in the introduction
to \chaud.

\bigskip\noindent{\it Components Of The Same Topological Type}

To complete and clarify the picture, and as preparation for the appendix,
we should perhaps add the following.
For bundles of a particular topological type on $\T^n$ with $n>2$,
there can be several different components of the moduli space
of flat connections and hence several different components of
the moduli space of superstring vacua. So the number of models
is more than one would get from the topological classification alone.
 
Here is a simple example for $n=3$ and a topologically trivial
bundle.  Consider a $\T^3/\Z_2$
orientifold with a $\Z_2$-invariant configuration of 32
sixbranes on $\T^3$, with an odd number of sixbranes at each of the
eight orientifold planes.  It can be shown that this is $T$-dual
to a superstring compactification on $\T^3$ with an $SO(32)$
bundle that obeys $w_1=w_2=w_3=0$.\foot{This was demonstrated
as follows by E. Sharpe.
In computing the mod two cohomology classes $w_1$ and $w_2$, pairs
of orientifolds will not contribute, so one can consider the case
of just eight sevenbranes, one at each fixed point.  This corresponds
to an $SO(8)$ bundle whose total Stieffel-Whitney class is
$\prod_{i=1}^8(1+x_i)$, where $x_i$ run over all eight elements
of $H^1(\T^3,\Z_2)$.  One can compute that this product equals 1,
modulo two and modulo four-dimensional classes (which vanish
as we are in three dimensions).  Likewise, an $SO(7)$ bundle that is the
direct sum of the seven non-trivial real line bundles of order two over
$\T^3$ has total Stieffel-Whitney class $\prod_{i=1}^7(1+x_i)=1$, where
now the $x_i$ run over the seven non-zero elements of $H^1(\T^3,\Z_2)$,
and is topologically trivial.  This assertion will be important in the
appendix.}
  These are sufficient conditions
for an $SO(32)$ bundle in three dimensions to be topologically
trivial.  So the model is connected topologically to
the standard $SO(32)$ model with the trivial flat connection
on a trivial bundle.  However, it is not connected to the trivial
flat connection via a family of flat connections; it is contained
in a component of the moduli space of flat connections in which there
are non-simply-laced unbroken
symmetry groups, such as $SO(25)$ (at level one),
while the trivial flat connection is in a component of the moduli
space of flat connections in which 
all unbroken symmetry groups are simply-laced.

There is no analog of this phenomenon on $\T^2$.  By using the
complex structure of $\T^2$, one can prove (as in a footnote 
in section 2.2), that for any given  semi-simple gauge group $G$
(regardless of whether $\pi_1(G)$ vanishes), the moduli spaces of
flat connections on a bundle of a given topological type are all connected
and irreducible.

For sufficiently large $n$, there are different components
of the moduli space of flat connections on a topologically non-trivial
$\Spin(32)/\Z_2$ bundle on $\T^n$, just as we saw above for topologically
trivial bundles on $\T^3$.  To construct such bundles
in the orientifold language, one simply places odd numbers of
branes at orientifold points of type ${\cal O}^+$ (the ones that
give $SO$ gauge symmetry) in such a way that $w_1=w_2=0$.
These last restrictions (which cannot be obeyed in the case 
$n=2$) are needed to get a $\Spin(32)/\Z_2$ model,
as opposed to an $O(32)/\Z_2$ or $SO(32)/\Z_2$ model that could
not make sense for the $\Spin(32)/\Z_2$ superstring.

\newsec{Threebrane Probes And $F$-Theory}

\subsec{Fivebranes And Classical Gauge Theory }

In this section, we consider  Type I fivebranes wrapped on $\T^2$
and their $T$-dual, which will involve threebrane probes of the
$\tilde \T^2/\Z_2$ orientifold.  By following a familiar logic
\refs{\sen,\probe},
we will deduce from the behavior of the threebrane probe an
$F$-theory description of the orientifold.

A system of $k$ parallel Type I fivebranes has a world-volume
gauge symmetry whose Lie algebra is that of $Sp(k)$
\ref\small{E. Witten, ``Small Instantons In String Theory,'' Nucl. Phys.
{\bf B460} (1996) 541.}.  
There are hypermultiplets that transform
under $Sp(k)\times SO(32)$ in the representation $\bf {2k}\otimes
{\bf 32}$, where the two factors are the basic representations of
$Sp(k)$ and $SO(32)$, respectively.  (There also are hypermultiplets
transforming in the traceless antisymmetric tensor of $Sp(k)$.)
In compactification without vector structure, there is a mod
two obstruction to the existence of particles transforming in
the $\bf{32}$ of $SO(32)$.  In order for the $\bf{2k}\otimes \bf {32}$ to
make sense, there must therefore be an equal (and canceling) obstruction
to the existence of the $\bf{2k}$.

In our problem of compactification on $\R^8\times\T^2$ without vector
structure, we wish to consider $k$ parallel fivebranes whose
world-volumes will be $\R^4\times \T^2$, where $\R^4$ is a subspace
of $\R^8$.  From what has been said above, the fivebrane world-volume
gauge group is effectively really $Sp(k)/\Z_2$, with no
``symplectic structure'' on the $\T^2$.  
Hence a supersymmetric state of the fivebrane system should be
described by a flat connection whose holonomies $B$ and $C$ (along
the two directions of $\T^2$) commute in $Sp(k)/\Z_2$, but if lifted
to $Sp(k)$ obey
\eqn\ikko{BC=-CB.}

The analysis described in section 2 for $\Spin(4n)/\Z_2$ flat connections
without vector structure has a very close analog for $Sp(k)/\Z_2$ flat
connections without symplectic structure.  As in section 2, an important
role is played by certain special examples of such flat connections
with maximal unbroken symmetry:

(1) Picking a subgroup $Sp(1)\times O(k)$ of $Sp(k)$ (under
which the $\bf{2k}$ of $Sp(k)$ transforms as $\2\otimes\bf k$), 
we take $B=b\times 1$, $C=c\times 1$, where
$b,c$ are elements of $SU(2)=Sp(1)$
with $bc=-cb$.  This gives a flat connection, unique up to gauge
transformation, with unbroken $O(k)$.

(2) For $k$ even, picking a subgroup $O(2)\times Sp(k/2)$ of
$Sp(k)$ (under which the $\bf {2k}$ of $Sp(k)$ transforms as 
$\2\otimes\bf k$), we take $B=b\times 1$, $C=c\times 1$, where
$b,c$ are elements of $O(2)$ with $bc=-cb$.  This gives a flat
connection with unbroken $Sp(k/2)$; as we saw in section 2, there
are three choices for $b,c$, up to conjugation.  If $k$ is odd,
a slight variant of this construction gives unbroken $Sp((k-1)/2)$.

Arguments just like those in section 2 for the $SO(4n)/\Z_2$ case
show (in keeping with the general theorem mentioned in a footnote
in section 2.2)
that the four flat connections just described are contained 
in one component of the moduli space of flat $Sp(k)/\Z_2$ connections
on $\T^2$.   Moreover, the moduli of these flat bundles can be
described in terms of the motion of a $\Z_2$-invariant
configuration of $k$ threebranes on an orientifold
$\tilde \T^2/\Z_2$.  For reasons very similar to those seen in section
2, the $\tilde \T^2$ that appears here
is  {\it not} the usual dual torus (which parametrizes
flat $U(1)$ connections on the original $\T^2$) but has half the size.
The case considered in (1) above is the case of $k$ threebranes
at the ${\cal O}^-$ orientifold plane, which we recall gives
$Sp$ gauge symmetry for sevenbranes but which evidently gives
$SO$ for threebranes.  Conversely, case (2) is the case of $k$
threebranes at one of the three ${\cal O}^+$ orientifold planes,
which give $SO$ gauge symmetry for sevenbranes but evidently give
$Sp$ gauge symmetry for threebranes.  
When $k$ is even, instead of speaking
of a $\Z_2$-invariant configuration of $k$ threebranes on $\tilde\T^2$,
we can speak of $k/2$ threebranes on $\tilde\T^2/\Z_2$.  When $k$ is odd,
there is a single threebrane ``stuck'' at the ${\cal O}^-$ point plus
$(k-1)/2$ pairs, so one cannot quite reduce the discussion to
branes on the quotient $\tilde \T^2/\Z_2$.
  The restriction to even $k$ in (2) above reflects
the fact that for odd $k$, one cannot move all threebranes to
a ${\cal O}^+$ point.

Instead of proving all of these statements by writing out formulas
similar to those in section 2, it seems more illuminating to give
direct arguments for the first cases $k=1,2$; these will actually
suffice for our applications.  For $k=1$, we have $Sp(1)=SU(2)$
and group elements $B, C$ with $BC=-CB$ are uniquely determined up to
conjugation.  This is the case of one threebrane stuck at ${\cal O}^-$,
with no moduli.

For $k=2$, which is the first instance in which one really
sees the full structure, we embed $Sp(1)\times SO(2)\subset Sp(2)$
(with as usual the ${\bf 4}$ of $Sp(2)$ transforming as $\2\times\2$)
and write $B=b\times b'$, $C=c\times c'$ (here $b,c\in Sp(1)$
and $b',c'\in SO(2)$), with $bc=-cb$ and
$b'c'=c'b'$, so $b$ and $c$ are uniquely determined up to conjugation and
\eqn\oppo{\eqalign{b'&=\left(\matrix{\cos\theta & \sin\theta\cr
            -\sin\theta & \cos\theta\cr}\right) \cr
c'&=\left(\matrix{\cos\psi & \sin\psi\cr
            -\sin\psi & \cos\psi\cr}\right), \cr}}
for some $\theta$, $\psi$.  This gives a surjective map from $SO(2)$
flat connections to $Sp(2)/\Z_2$ flat connections without symplectic
structure, but there are some discrete identifications, as in section 2.
To see these and to elucidate the structure of the moduli space,  we
use the fact that $Sp(2)/\Z_2=SO(5)$.  In $SO(5)$, $B$ and $C$ become
in a suitable basis
the commuting matrices
\eqn\pnoppo{\eqalign{B &=\left(\matrix{-1 &0 &0 &0 &0 \cr 
             0& 1 &0 &0 &0 \cr
             0 &0 & -1 &0 &0 \cr
              0 &0&0&\cos 2\theta & \sin 2\theta \cr
                 0&0&0& -\sin 2\theta & \cos 2\theta\cr           }\right)\cr
                 C &=\left(\matrix{1 & 0&0 &0 &0 \cr 
            0 & -1 &0 &0 &0 \cr
             0 &0 & -1 &0 &0 \cr
              0 &0&0& \cos 2\psi & \sin 2\psi \cr
                 0&0&0& -\sin 2\psi & \cos 2\psi\cr           }\right).\cr}}
The fact that these formulas depend trigonometrically on $2\theta$
and $2\psi$ shows that there are symmetries $\theta\to\theta+\pi$
and (independently) $\psi\to\psi+\pi$, as a result of which the torus
$\tilde \T^2$ appearing in the orientifold has one-half the usual size.
From \pnoppo, it is straightforward to see that the generic unbroken
gauge symmetry is $U(1)$,\foot{Just as in section 2, we will somewhat
imprecisely use the name ``unbroken gauge symmetry'' to refer to the subgroup
of $Sp(2)/\Z_2$ consisting of group elements which if lifted to $Sp(2)$
commute with $B$ and $C$.  Including also those that anticommute with
$B$ or $C$ gives  an additional $\Z_2\times \Z_2$
 generated by matrices ${\rm diag}(\pm 1,\pm 1, \pm 1, 1,1)$ of
determinant one.
These are present for all values of $\theta,\psi$, commute with $U(1)$
and its extensions described below, and are related
to the fact that the winding lattice of the dual torus has one-half
the usual size.}
 and that the unbroken symmetry is
enhanced in precisely the following two cases:
$(1')$ If $\theta=\psi=0$, an extra $\Z_2$ appears (generated
by the matrix ${\rm diag}(-1,-1,-1,-1,1)$) and the unbroken
symmetry is $O(2)$ rather than $SO(2)=U(1)$.  
$(2')$ If $2\theta$ and $2\psi$ are both 0 or $\pi$ (and not both
zero) there is an unbroken $SO(3)=Sp(1)/\Z_2$. (For example,
if $2\psi$ and $2\theta$ are both $\pi$, then the unbroken $SO(3)$
acts in the $3\cdot4\cdot 5$ subspace in the basis used in \pnoppo.)
These two cases are just the specialization to $k=2$ of
the special cases (1), (2) of $Sp(k)/\Z_2$ flat
connections listed earlier.  The modulus
$a=(2\theta,2\psi)$ up to the $SO(5)$ Weyl transformation $a\to -a$
parametrizes the position of a threebrane on the orientifold 
$\tilde \T^2/\Z_2$.

The fact that for $k=2$ -- two fivebranes on $\T^2$ -- we get a dual
description with only {\it one} threebrane on $\tilde \T^2/\Z_2$
is, of course, another manifestation of the curious factors of two
that have appeared throughout this paper.

\subsec{$F$-Theory Interpretation}

Now our goal is to give an $F$-theory interpretation of $\T^2$
compactification without vector structure.  At first sight, this poses
a conundrum, since the $F$-theory description must involve
$F$-theory on a Calabi-Yau two-fold, but there is no obvious
suitable candidate.  $F$-theory on K3 is the heterotic string on $\T^2$
with vector structure, and $F$-theory on $\T^2\times \T^2$ is simply
Type IIB on $\T^2$.

The answer that we will find is that $\T^2$ compactification without
vector structure is dual to $F$-theory on a K3 which is constrained
to have a novel kind of
$D_8$ singularity.  This $D_8$ singularity looks macroscopically
like an ordinary $D_8$ singularity (at long distances from it there
is a standard $D_8$ ALE space), but has in a
mysterious way absorbed some sort of flux, such that it cannot
be blown up or deformed away and does not generate gauge symmetry.
If however this $D_8$ is extended to a $D_{8+m}$ singularity, then
 $Sp(m)$ gauge symmetry appears.

To obtain this result, we simply examine as in \probe\ the theory
on a threebrane probe.
We consider as in section 4.1 a pair of fivebranes wrapped on $\T^2$,
dual to a single threebrane probe on $\tilde \T^2/\Z_2$.
The ``base space'' $B=\tilde \T^2/\Z_2$ is as a complex manifold
isomorphic to ${\bf CP}^1$.  On the threebrane probe, there is a $U(1)$
gauge field whose $\tau$ parameter determines up to isomorphism
an elliptic curve that varies holomorphically with the position
of the probe on $B$.  By determining the structure of the resulting
elliptic fibration $X\to B$, one obtains an $F$-theory description.
As already suggested, $X$ will be a K3 surface with a ``frozen''
$D_8$ singularity.

The underlying theory of two fivebranes and 32 ninebranes
has gauge symmetry $Sp(2)\times SO(32)$,
with hypermultiplets transforming as $\bf 4\otimes {\bf 32}$
(the tensor product of the fundamental representations of the two
groups) plus $\bf 5\otimes 1$, where the $\bf 5$ is the traceless
antisymmetric tensor of $Sp(2)$ or equivalently the vector of $SO(5)$.

In the dual description on $\tilde \T^2/\Z_2$, 
for a generic threebrane position, the unbroken gauge symmetry
is a $U(1)$ that appears as the second factor
in $Sp(1)\times U(1)\subset Sp(1)\times O(2)
\subset Sp(2)$. (Here as before, the ${\bf 4}$ of $Sp(2)$ is
$\2\otimes\2$ of $Sp(1)\times O(2)$, fixing the embeddings in the
chain just mentioned.)  In the $\bf 4$ of $Sp(2)$, the unbroken $U(1)$
has charges $1$ and $-1$, while in the $\bf 5$ (as it arises in the
tensor product of two $\bf 4$'s) the charges are $2$, $-2$, and $0$.
The factor of 2 in the charges of the $\bf 5$ relative to the $\bf 4$
will be important.  (This factor has actually already appeared in \pnoppo.)

The details of the elliptic fibration over $B$ depend on where we place
the eight sevenbranes.  We place them in generic, distinct points.
Now in the motion of the threebrane, there are 12 exceptional cases
at which monodromy occurs.  The threebrane may collide with one of the
eight sevenbranes, or with one of the four orientifold planes.  There are
three essential cases:

(1) If the threebrane meets a sevenbrane, a single charge one hypermultiplet
becomes massless. (In the Type I description, it originates in the
${\bf 4}\otimes {\bf 32}$, which is why it has charge 1.)
This gives according to \ref\sw{N. Seiberg and E. Witten,
``Electric-Magnetic Duality, Monopole Condensation, and Confinement
In $N=2$ Supersymmetric Yang-Mills Theory,'' Nucl. Phys. {\bf B426} (1994) 19,
hep-th/9407087, ``Monopoles, Duality, and Chiral Symmetry Breaking
In $N=2$ Supersymmetric QCD,'' hep-th/9408099.} a monodromy conjugate to
\eqn\ubbbu{T=\left(\matrix{ 1 & 1 \cr 0 & 1\cr}\right).}
In all we get eight such points from the eight sevenbranes.

(2) If the threebrane meets an orientifold plane of type ${\cal O}^+$,
one gets at the classical level an enhanced $SO(3)$ gauge symmetry
with no massless hypermultiplets.  At the quantum level \sw, this
splits into a pair of quantum singularities of monodromy $T$.
From the three ${\cal O}^+$ planes, we get $2\times 3 = 6$ such points.

(3) The remaining case is that the threebrane meets an orientifold
plane of type ${\cal O}^-$.  In this case, at the classical level, 
the $U(1)$ gauge symmetry of the threebrane is enhanced to $O(2)$.
Two hypermultiplets, of charge 2, become massless at this point.
They are components of the 
${\bf 5}$ of $SO(5)$ that correspond to the two eigenvalues $B=C=1$ that
can be seen in 
eqn. \pnoppo\ at $\theta=\psi=0$, and have charge two because, as explained
above, the charged components of the $\bf 5$ have charge $\pm 2$
(or equivalently, because of the factor of 2 multiplying $\theta$ and
$\psi$ in \pnoppo).
These give a monodromy conjugate to $-T^4$, as we will compute momentarily.

To verify the  assertion that the monodromy is $-T^4$, 
we note that in $U(1)$ gauge theory
with several hypermultiplets of charge $q_i$ and zero bare mass, 
the monodromy around the 
 singular point at the origin of the Coulomb branch
(where $a=a_D=0$ and the hypermultiplets become massless) 
is conjugate to $T^\Delta$ with $\Delta=\sum_iq_i^2$.
In the present case, with two hypermultiplets of charge 2, this would
give $\Delta=8$.  However, we actually  have $O(2)$ and not $SO(2)$
gauge symmetry,
which simply means that we must divide by a discrete transformation
that maps $(a,a_D)\to (-a,-a_D)$.  This gives an extra factor
of $-1$ in the monodromy.  In addition, since the monodromy we want
corresponds to going only ``half-way'' around the origin in the complex
$a$-plane, the exponent $\Delta$ is replaced by $\Delta/2$.  So the monodromy
for gauge group $O(2)$ is not conjugate to
$T^8$, as it would be for $U(1)$, but to $-T^4$.

An elliptic fibration with section that has a singular fiber of monodromy
conjugate to $-T^4$ has, in its Weierstrass model, a singularity of
type $D_8$ on that fiber.  So the elliptic fibration $X\to B$ that
describes $\T^2$ compactification without vector structure has
a $D_8$ singularity.  It also has, counting the results of (1) and (2)
above, $8+6=14$ fibers with generic singularities (nodes or ordinary
double points) with monodromy conjugate to $T$.  Ordinarily, a $D_8$
singular fiber can be deformed to 10 generic singular fibers,
which in the present context would give  an elliptic fibration $X\to B$ 
with $14+10=24$ generic singular fibers.  Such an $X$ would have
Euler characteristic 24, and would be a K3 surface.  In the present
context, evidently, as there are only eight sevenbranes, one is not
permitted to deform away the $D_8$ singularity.  Evidently,
$\T^2$ compactification without vector structure is dual to 
$F$-theory on an elliptic  K3 surface $X$ with
an irremovable $D_8$ singularity.

If $m$ of the $8$ sevenbranes on $B={\bf P}^1$ approach the ${\cal O}^-$ point,
then according to the analysis in section 2, we get an enhanced
$Sp(m)$ gauge symmetry, at level one.  On the other hand, moving
$m$ sevenbranes to a $D_8$ fiber enhances the singularity to $D_{8+m}$.
Hence, we conclude that an $F$-theory $D_8$ singularity of this type
has the property that if it is enhanced to $D_{8+m}$, a level
one $Sp(m)$ gauge symmetry is generated in spacetime.

\subsec{Reduction To $M$-Theory}

Now we would like to discuss what happens when one compactifies
further, to go to an $M$-theory description.

Since Type I on $\T^2$ without vector structure is equivalent to
$F$-theory on a K3 surface $X$ with a $D_8$ singularity,
one might think that Type I on $\S^1\times \T^2$, with an obstruction
to vector structure on the $\T^2$, would be equivalent to $M$-theory
on the same surface $X$.   It seems, however, that this is not quite
so.

If we start with Type I on $\S^1\times \T^2=\T^3$, and perform
$T$-duality, we get an orientifold $(\tilde \S^1\times \tilde \T^2)/\Z_2$
with eight orientifold sixplanes and eight sixbrane pairs.
(The factors $\tilde \S^1$ and $\tilde \T^2$
are of course the duals to the factors $\S^1$ and $\T^2$ in the original
$\T^3$).  Two of the orientifold planes are of type ${\cal O}^-$,
and six are of type ${\cal O}^+$.  
A quick way to verify these statements is to look at the maximal
unbroken symmetry groups.  On $\T^2$ without vector structure, a maximal
symmetry group was $Sp(8)$ and it occurred in a unique fashion up to
gauge transformation.  On $\S^1\times \T^2$, with an obstruction to
vector structure that comes from the second factor, $Sp(8)$ is still
a maximal symmetry group, but it occurs in {\it two} ways up to
gauge transformation. (1) One can ``pull back'' a flat connection
with $Sp(8)$ symmetry from the second factor in $\S^1\times \T^2$.
(2) One can also ``twist'' it by a Wilson line on the $\S^1$ factor
with holonomy in the center of $Sp(8)$; as this center is $\Z_2$,
this gives precisely one new possibility.  Likewise, the second
maximal symmetry group, which is $O(16)$, appears three times
on $\T^2$ but six times on $\S^1\times \T^2$.  This counting of
flat connections with maximal symmetry shows that the numbers of
${\cal O}^-$ and ${\cal O}^+$ sixplanes are two and six.  The rank of
the possible symmetry groups shows that the number of sixbrane pairs
is eight.  One can also verify vanishing of the net sixbrane charge.
The two kinds of orientifold plane have charge $+4$ and $-4$, and
one has $2\cdot 4-6\cdot 4+2\cdot 8=0$.

\nref\seibergs{N. Seiberg, ``IR Dynamics On Branes And Space-Time Geometry,''
Phys. Lett. {\bf B384} (1996) 81, hep-th/9606017.}
\nref\seiwi{N. Seiberg and E. Witten, ``Gauge Dynamics And Compactification
To
Three Dimensions,'' in {\it The Mathematical Beauty Of Physics}, eds.
J. M. Drouffe and J. B. Zuber (World Scientific, 1997) p. 333, hep-th/9607163.}
\nref\moresen{A. Sen, ``A Note On Enhanced Gauge Symmetries In
$M$ And String Theory,'' hep-th/9707123.}
So this model corresponds to a Type IIA orientifold on $(\tilde \T^3/\Z_2)$
with two ${\cal O}^-$ sixplanes and six ${\cal O}^+$ sixplanes.
We expect the model to have an $M$-theory description in terms of
compactification on a (perhaps singular)
K3 surface because it was obtained by compactifying
on an extra circle from a model that had such an $F$-theory description.
In $M$-theory, a sixplane of type ${\cal O}^+$ does not correspond
to any singularity \refs{\seibergs - \moresen}.  To account for
the fact that $\S^1\times \T^2$ compactification of Type I
without vector structure gives eight fewer vector multiplets
than the usual case, each of the two ${\cal O}^-$ planes must
lead to a singularity of the K3 surface that reduces the rank of the 
gauge group by four.  By analogy with what we found in the $F$-theory case,
the most obvious possibility is that an ${\cal O}^-$ sixplane is
converted in $M$-theory to a rank four $A-D-E$ singularity that does
not produce gauge symmetry.

In fact, by showing that this hypothesis makes it possible to solve certain
four-dimensional gauge theories, Landsteiner and Lopez
\ll\ argued fairly convincingly that an ${\cal O}^-$ sixplane
corresponds to a $D_4$ singularity which does not generate gauge symmetry.
(They expressed the singularity associated with the
${\cal O}^-$ sixplane as the quotient
of the singular surface $xy=v^4$ by a $\Z_2 $ symmetry
$x\leftrightarrow y, \,v\leftrightarrow -v$.  By introducing the
invariant functions $w=v^2$, $a=(x-y)v$, $b=x+y$, which obey
$a^2=wb^2-4w^3$, one sees that the quotient is in fact a $D_4$ singularity.)

We now have what at first looks like a contradiction between the following
facts.  (1) Type I on $\T^2$ without vector structure gives 
$F$-theory on a K3 surface $X$ with a $D_8$ singularity that does
not give gauge symmetry.
(2) $F$-theory on $\S^1\times X$, for any $X$, is equivalent to $M$-theory
on $X$.  (3) Type I on $\S^1\times \T^2$, with an obstruction to vector
structure coming from the $\T^2$, is equivalent to $M$-theory on
$X$ with two $D_4$ singularities that do not give gauge symmetry.
The apparent contradiction is that according 
to (1), it seems that the Type I model on $\S^1\times \T^2$ under
consideration here should give $M$-theory on a K3 with a $D_8$ singularity,
but according to (3), the actual singularity is $D_4\times D_4$.

The resolution of the issue apparently has to do with the following
subtlety about $F$-theory, which affects the precise meaning of statement
(2).  In $F$-theory, one really only has a variable
$\tau$ parameter, with certain singularities, over a base $B$.  One
then interprets this geometrically in terms of an elliptic fibration
given in a Weierstrass model.  (Given the $\tau$ parameter as a function
on $B$, one has the data to canonically construct a Weierstrass model
by writing the equation $y^2=4x^3-g_2(\tau)x-g_3(\tau)$ which can
be written once $\tau$ is known.  One does not generally have the
information to canonically write other models, but under certain
conditions, for instance if $X$ is singular, other models obeying
a Calabi-Yau condition do exist.) 
For example, when we assert that Type I on $\T^2$, without vector
structure, is equivalent to $F$-theory on a K3 surface with a $D_8$
singularity, we really mean that the monodromy of the $\tau$ parameter
is such that the Weierstrass model of the corresponding elliptic fibration
has a $D_8$ singularity.

After compactifying on an additional circle and going to $M$-theory, 
the fibers of the elliptic fibration become ``physical'' and one
gets an actual geometrical K3 surface.  Evidently, in the case
of $\T^2$ compactification without vector structure, the model
that results after compactifying on an additional circle is not the
Weierstrass model, with its $D_8$ singularity, but a different model
of the same elliptic fibration, with two $D_4$ singularities.
I will not explain why this is so, but I will show that the second
model does exist.  

By blowing up a $D_8$ singularity in a complex surface, one gets a 
configuration
of eight genus zero curves arranged according to the Dynkin diagram
of $D_8$.  (Each of the eight genus zero curves corresponds to a node
in the $D_8$ Dynkin diagram, and two nodes in the Dynkin diagram are
connected if and only if the corresponding curves intersect.)  If
the $D_8$ singularity is contained in a fiber of an elliptic fibration,
then the fiber itself is a ninth genus zero curve (the generic fiber
of an elliptic
 fibration has genus one, but fibers containing singularities
have genus zero).  The nine curves are arranged according to the
Dynkin diagram of $\widehat D_8$ (the affine extension of $D_8$).

Given a K3 surface that contains a configuration of nine genus zero
curves arranged according to the $\widehat D_8$ Dynkin diagram, one can
produce a variety of birational models (which are all singular
Calabi-Yau manifolds) by ``blowing down'' some of them
to produce a singularity.  The blown down curves make up a subdiagram
of the $\widehat D_8$ diagram, and this subdiagram determines the type of
singularity.  For instance, in the Weierstrass model, eight curves
forming a $D_8$ subdiagram of the $\widehat D_8$ diagram
are blown down, giving
a $D_8$ singularity. This is the model of  a singular K3 surface $X$ that
can be used for an $F$-theory description of Type I on $\T^2$ without
vector structure.

In general, one can blow down any proper subset of the nine curves
(but not all nine) to get a singular Calabi-Yau manifold
 that is birational to $X$.  
For example, if one omits the ``middle'' node
of the $\widehat D_8$ diagram, one is left with the Dynkin diagram
of $D_4\times D_4$.  By blowing up the curve that corresponds to the 
``middle'' node, and blowing down all others, one gets a K3 surface $Y$
that is birational to $X$ and has two $D_4$ singularities.  From what
has been said above, $M$ theory on $Y$ is equivalent to Type I
on $\S^1\times \T^2$ without vector structure.  The $D_4$ singularities
have of course, like the $D_8$ singularity in the analogous $F$-theory
version, the property that they do not generate gauge symmetry and
cannot be resolved or deformed away.  

\newsec{Duality To CHL Model}

$\T^2$ compactification without vector structure has at least one more
dual description.  It is equivalent to a form of the CHL model,
in fact, to compactification of the
$E_8\times E_8$ heterotic string on $\T^2$ with the two $E_8$'s
swapped in going around one circle of the $\T^2$.  This has
been somewhat implicitly sketched in \chaud, and derived more fully
in \lerche.  
After giving a direct explanation of why this is true, we will give
a more theoretical explanation, which will also show that within a certain
class of constructions, there are no more models of the same kind.

In fact, we can construct either the CHL model or $\T^2$ compactification
without vector structure via an involution (that is, a $\Z_2$
symmetry) of a Narain lattice $\Gamma^{17,1}$, as follows:

$(A)$ We write $\T^2=\S^1\times \S^1$ and call the two factors
the ``first'' and ``second'' $\S^1$, respectively.  To build
the CHL model, we consider conventional $\S^1$ compactification
of the $E_8\times E_8$ heterotic string.  This is described by
a Narain lattice $\Gamma^{17,1}$, with a particular decomposition
into left- and right-movers.   Picking a decomposition
$\Gamma^{17,1}=\Gamma^8\oplus \Gamma^8\oplus \Gamma^{1,1}$,
where the $\Gamma^8$'s are copies of the root lattice of $E_8$,
we define an involution $w$ of $\Gamma^{17,1}$ that exchanges
the two $\Gamma^8$'s (breaking $E_8\times E_8$ to a diagonal $E_8$)
and acts trivially on $\Gamma^{1,1}$.  To get the
CHL model, we compactify on the second circle, with a monodromy
around the second circle consisting of the automorphism $w$ of
the Narain lattice.

\def\tilde{\widetilde}
$(B)$ We pick $\Spin(32)/\Z_2$ matrices $U,V$ with $UV=-VU$.
We compactify the $\Spin(32)/\Z_2$ heterotic string
(which of course is equivalent to the Type I heterotic string)
on the first circle with a Wilson line $U$.  This gives a model
that can be described by a Narain lattice $\Gamma^{17,1}$, with a particular
decomposition into left- and right-movers, and which is in particular
equivalent (up to motion in moduli space) to the first step in $(A)$ above. 
The $\Spin(32)/\Z_2$ matrix $V$ gives an automorphism of this
theory which is an involution $\tilde w$ of the Narain lattice. 
Now, to obtain $\T^2$ compactification without vector structure,
we compactify on the second circle in $\T^2$, with a monodromy
around the second circle consisting of the automorphism $\tilde w$ of the
Narain lattice.

Note, that, if we wish, we can in either $(A)$ or $(B)$ take Wilson
lines along the first circle (corresponding in case $(A)$ to Wilson
lines in the diagonal $E_8$ and in case $(B)$ to a suitable
choice of $U$) for which the unbroken gauge group is  abelian.
In this case, there is no unbroken $SU(2)$ and the vacuum is determined
by a vector $v$ that does not lie on any of the walls introduced below.
This fact will be important later.
Also, the ability to pick convenient Wilson lines makes possible
the following very direct explanation of why the two models are the same.

In $(A)$, we start with a large radius $R$ on the first circle,
and a Wilson line on it that breaks $E_8\times E_8$ to $SO(16)\times SO(16)$.
There is of course a monodromy around the second circle that exchanges the
two $E_8$'s and therefore the two $SO(16)$'s.  Now we interpolate to
small $R$.  The $E_8\times E_8$ heterotic string, in its vacuum
with unbroken $SO(16)\times SO(16)$, on a circle of radius $R$,
is $T$-dual to   the $\Spin(32)/\Z_2$ heterotic
string, on a circle of radius $1/R$, likewise in a vacuum with unbroken
$SO(16)\times SO(16)$.  Of course, if $R$ is small,
the $\Spin(32)/\Z_2$ description
is more useful.  In this description, a Wilson line around the first circle 
breaks
$\Spin(32)/\Z_2$  to $SO(16)\times SO(16)$.  The  Wilson 
line in question is our friend $U=u\times 1$,
where $u$ is the $O(2)$ matrix introduced in eqn. \nokko.  The monodromy
around the second circle that exchanges the two $SO(16)$'s becomes in 
the $\Spin(32)/\Z_2$ description a Wilson line $V=v\times 1$, with 
$v$ as in \nokko.  So we have arrived at the familiar Wilson lines
$U=u\times 1$, $V=v\times 1$, on $\T^2$, showing that model $(A)$ is
$T$-dual to model $(B)$.

The relation of the $E_8\times E_8$ heterotic string to $M$-theory gives
a natural explanation for this result; we will explain it briefly.
  The $E_8\times E_8$ model
in ten dimensions is equivalent to $M$-theory on an interval $\S^1/\Z_2$.
Model $(A)$ is obtained by compactifying further on $\S^1\times \S^1$
in such a way that in going around the first $\S^1$, the two
end-points of $\S^1/\Z_2$ are exchanged.  We write the resulting three-manifold
with boundary
as $(\S^1/\Z_2\times_W\S^1)\times \S^1$, where $W$ is the ``flip'' that
exchanges the two ends of $\S^1/\Z_2$.  (Thus, $\S^1/\Z_2\times_W\S_1$ is
a M\"obius strip; the equivalence of the CHL  model to $M$-theory on a M\"obius
strip has been noted in 
\nref\park{A. Dabholkar and J. Park, ``Strings On Orientifolds,''
hep-th/9604178.}\nref\gp{C. V. Johnson, private communication.}
\refs{\park,\gp}.)
On the other hand, $M$-theory on $(\S^1/\Z_2\times_W\S^1)\times \S^1$
is equivalent to Type IIA on $\S^1/\Z_2\times_W\S^1$.  
We note now that Type IIA on $\S^1/\Z_2$ should be interpreted 
\refs{\dlp,\horava} as a $T$-dual of Type I on $\S^1$
with $SO(32)$ broken to $SO(16)\times SO(16)$ and
an unbroken $SO(16)$ at each fixed point;
in this Type I description, $W$ becomes a gauge transformation that exchanges
the two $SO(16)$'s.  Fibering 
now over $\S^1$ with monodromy $W$ generates the familiar
Type I model without vector structure on $\S^1\times \S^1$.

\bigskip\noindent{\it Geometry Of The $T$-Duality Group}

We now want to show that, in a sense, models $(A)$ and $(B)$ ``must'' be
the same, since there is only one model of their kind.
For this, we must explain something about the geometry
of $T$-duality for the heterotic string in nine dimensions.
The duality group is not quite the full symmetry group $O(17,1;\Z)$
of the lattice $\Gamma^{17,1}$, because one is limited to transformations
that do not reverse the direction of ``time,'' or in other words
do not change the sign of the right-moving momentum.  (Transformations
that reverse the sign of the right-moving momentum do not respect the
GSO projection of the heterotic string.)   The subgroup
of $O(17,1;\Z)$ that preserves the direction of time will be called
$F=O_+(17,1;\Z)$.  The Narain moduli space $\CM=
O_+(17,1;\Z)\backslash O_+(17,1;\R)/O(17;\R)$
can be constructed in two steps.   The quotient
$H=O_+(17,1;\R)/O(17;\R)$ is the hyperboloid that parametrizes timelike
future-pointing
vectors $v$ of length squared $-1$ in $\R^{17,1}$.  Physically, in the
application to the heterotic string, $v$ generates the subspace of $\R^{17,1}$
that (at a given point in the moduli space) consists of right-moving momenta.
The moduli space $\CM$ of vacua is the quotient $F\backslash H$.  So it is 
useful to understand the action of $F$ on $H$.

For every vector $m\in \Gamma^{17,1}$ of length squared 2, there is
an elementary reflection $R_m$ in $O_+(17,1;\Z)$, which acts by
$v\to v-m(m,v)$.  The condition for $v$ to be a fixed point of this
transformation is $(m,v)=0$, which means that in the heterotic
string vacuum determined by $v$,  $m$ is purely left-moving
and so is the highest root of an enhanced $SU(2)$ gauge symmetry
(in which $R_m$ is a Weyl transformation), which is unbroken when
$(m,v)=0$.
The condition $(m,v)=0$ defines a hyperplane $T_m$ in $H$ which we will
call a wall.

According to theorem 1 of \ref\conway{J. H. Conway, ``The Automorphism
Group Of The 26-Dimensional Even Unimodular Lorentzian Lattice,''
in J. H. Conway and N. J. A. Sloane, {\it Sphere Packings, Lattices, and
Groups} (Springer-Verlag, 1988), p. 527. }, the subgroup $F'$ of
$F$ that is generated by the reflections $R_m$ for various $m$ is
of index two.  The $F'$ action on $H$ can be fairly conveniently described.
The walls $T_m$ divide $H$ into regions which (as can be shown via
general arguments about reflection groups) are fundamental domains
for the action of $F'$.  Each such fundamental domain is bounded
by nineteen walls, which correspond to points $m_i$, $i=1,\dots, 19$,
of length squared two.  (Each fundamental domain
goes to infinity in two ways, corresponding to unbroken $E_8\times E_8$ or
$\Spin(32)/\Z_2$.) The $m_i$  are called the simple positive roots
of $\Gamma^{17,1}$ (for the given choice of fundamental domain).  Their
matrix of inner products determines a Dynkin (or Coxeter)
diagram, with 19 nodes,
that is associated with the lattice $\Gamma^{17,1}$.  It is drawn on
p. 529 of \conway, and looks like two copies of the extended Dynkin
diagram of $E_8$ (a total of $2\times 9 = 18$ nodes) combined by
attaching the extended nodes at the end of the $\widehat E_8$ diagrams
to a nineteenth node.  This Dynkin diagram has precisely one diagram
automorphism $\Upsilon$, 
which exchanges the two $E_8$'s.\foot{The same diagram automorphism
also acts as a diagram automorphism of a $\widehat D_{16}$ diagram
that is a subdiagram of the $\Gamma^{17,1}$ diagram.  The fact that
$\Upsilon$ induces diagram automorphisms of either $\widehat E_8\times \widehat
E_8$ or $\widehat D_{16}$ subdiagrams can possibly serve as the starting
point for an alternative explanation of the relation 
between constructions $(A)$ and $(B)$.} 
This diagram automorphism
generates  the group $F/F'\cong \Z_2$ of symmetries of $\Gamma^{17,1}$
that cannot be obtained as products of reflections.  As we will
see, it is responsible for the existence of models $(A)$ and $(B)$.   

In building models along the lines of $(A)$ or $(B)$, the idea is
to first compactify on a circle $\S^1$ with a vacuum determined by some
point in $H$ that is invariant under some $x\in F$.  Because of this
$H$-invariance, it is possible, in compactifying on a second circle,
to obtain a ``twisted'' model by saying that the physics is rotated
by $x$ in going around the second circle.  So in analyzing the possible
monodromies of twisted models, we can restrict ourselves to $x$'s that
have fixed points in $H$.

A case that (as suggested by J. H. Conway) is quite pertinent
and can be analyzed very quickly
is the case that $x$ leaves fixed a point in the interior of a fundamental
domain ${\cal F}$.  In this case, $x$ clearly permutes among themselves
the 19 boundary ``walls'' of ${\cal F}$, and therefore permutes the
19 simple roots $m_i$ that determine these walls.  Since $x$ is an element
of $O_+(17,1;\Z)$, in permuting the $m_i$ it preserves the geometric
relations among them.  So $x$ acts as an automorphism of the Dynkin
diagram, which, if not the identity, must be $\Upsilon$.  The $m_i$
generate $\Gamma^{17,1}$, so $x$ is determined by its action on them.
There is therefore (other than the identity) only one element of $F$
with a fixed point in the interior of ${\cal F}$, namely $\Upsilon$.

This suffices for comparing models $(A)$ and $(B)$, since (as noted earlier)
 both models are
constructed using monodromies that leave invariant  points in $H$ at
which the unbroken gauge group is abelian.  Such points do not lie on
any wall but rather lie
in the interior of a fundamental domain.  For a given fundamental domain,
there is only one choice of an element of $F$ (other than the identity)
that leaves fixed a point in its interior.  So, after conjugating by
an element of $F'$ that identifies the fundamental domains used in $(A)$
and $(B)$, these models are constructed with the same monodromy element
and hence coincide.

Furthermore, because of the uniqueness of $\Upsilon$, this is the only
model that can be built using a monodromy that leaves fixed a vacuum on the
first circle in
which the unbroken gauge group is abelian.  Can any new models
be constructed using monodromies that leave fixed only vacua with
nonabelian gauge symmetry? 
By arguments similar to the ones used to
show that the walls  divide $H$ into fundamental domains, it can
be shown that any model with monodromy in $F'$ is equivalent to standard
toroidal compactification.  (One shows that if an element $x\in F'$ leaves
fixed a vacuum corresponding to a vector $v\in H$, then $x$ is an element
of the Weyl group of the unbroken gauge group $G$ of this vacuum.  The
monodromy $x$ can be continuously rotated to the identity in $G$, showing
that the model twisted by $x$  is equivalent 
to standard $\T^2$ compactification.)
This leaves the question, which will not be resolved here, of whether
new models can be constructed using a monodromy $x$ that is not in $F'$
and leaves fixed only vacua with nonabelian gauge symmetry.  (Considerations
of supergravity show that, if so, the twisting by $x$ in going around
the second circle breaks the nonabelian
gauge symmetry, which will therefore not actually be present
in eight dimensions.)  

\newsec{The Third  Orientifold}

\subsec{$F$-Theory Description And $T$-Duality}

Apart from the most familiar
 case discussed in \refs{\dlp,\horava} and the additional 
case that we have already considered, there is one more supersymmetric
orientifold with target $\tilde \T^2/\Z_2$.  
Any such orientifold has $n_+$ sevenplanes of type ${\cal O}^+$, and
$n_-$ of type ${\cal O}^-$, with $n_++n_-=4$.  The models with
$(n_+,n_-)=(4,0)$ and $(3,1)$ are by now familiar; what remains
 is the model with $(n_+,n_-)=(2,2)$.
(A supersymmetric orientifold on $\tilde \T^2/\Z_2$ cannot have $n_->2$, 
because in that case the net sevenbrane
charge of the orientifolds would be positive, and could not be canceled
except by adding anti-sevenbranes and violating supersymmetry.)
The main
purpose of the present section is to understand some dual descriptions
of the $(n_+,n_-)=(2,2)$ model.

\bigskip\noindent{\it $F$-Theory Dual}

In fact, we can immediately identify an $F$-theory dual.  It will
be a K3 elliptic fibration over $\tilde\T^2/\Z_2\cong {\bf P}^1$ with
the following structure.  An
${\cal O}^+$ sevenplane splits nonperturbatively into
a pair of fibers with generic (nodal or normal crossing) singularities
\sen, and we saw in section 4 that an ${\cal  O}^-$ sevenplane 
is represented in $F$-theory as a $D_8$ singularity that, for presently
obscure reasons, cannot be resolved or deformed.  So the $(n_+,n_-)=(2,2)$
model has an $F$-theory description
via a K3 fibration over ${\bf P}^1$ with
two of these exotic $D_8$ singularities and four generic singular fibers.

\bigskip\noindent{\it Search For A $T$-Dual}

Our remaining attention will therefore be directed at seeking a $T$-dual
of the $(n_+,n_-)=(2,2)$ model.  It  is most natural, however, to first
consider a similar question that arises in compactification to {\it nine}
dimensions.  There are two Type IIA orientifolds
on $\S^1/\Z_2$.  One model, with
$(n_+,n_-)=(2,0)$, and 16 eightbranes (or 16 eightbrane pairs on the 
covering space $\S^1$), has been extensively studied.  The second
possibility is $(n_+,n_-)=(1,1)$.  In this case, the net eightbrane charge
of the orientifolds vanishes, so there are no eightbranes.
We will seek a $T$-dual of this model, that governs its behavior in the
limit that the radius of the $\S^1/\Z_2$ vanishes.  In seeking
this $T$-dual, we will be guided by the following clues:

(1) The ${\bf {RP}}^2$ contribution to the vacuum amplitude
vanishes, because of a cancellation
between the contributions from the two orientifold planes.  In
fact, there is a symmetry of the free closed string sector 
 -- the one that flips the two ends of $\S^1/\Z_2$ -- under
which the ${\bf {RP}}^2$ path integral is odd.  However,
there are unoriented world-sheets in the theory, and amplitudes
that violate the symmetry just mentioned
receive non-zero contributions from ${\bf {RP}}^2$.  For future
use, note that if $\S^1$ is parametrized by an angle $\theta$ with
$0\leq \theta\leq 2\pi$, then the symmetry that exchanges the two ends
of $\S^1/\Z_2$ is $\theta\to\theta+\pi$, which acts on a state of momentum
$p$ along the circle as $(-1)^p$; so the ${\bf {RP}}^2$ amplitude violates
conservation of $(-1)^p$.

(2) In the weak coupling limit, the spectrum is the same as that of the closed
string sector of the $(1,1)$ orientifold on $\S^1/\Z_2$, 
but the interactions are different.
By $T$-duality, the spectrum is also the same, in the weak coupling
limit, as that of a Type I superstring on $\S^1$.
This strongly suggests that the target space of the $T$-dual of
the $(1,1)$ orientifold should be 
$\S^1$; its spectrum on  an $\S^1$ of large radius should
match the spectrum of the $(1,1)$ orientifold for small radius.

\subsec{The $T$-Dual}

We will now describe a simple model that obeys the required properties.
\foot{This model was described in a different but equivalent
language, along with many of the properties discussed below, in section 3.1 of
\park.
  In a notation we will introduce presently, the model
was described there as a Type IIB orientifold with $M=\R^9\times \S^1$
and $\tau_M$ a $\pi$ rotation of $\S^1$.}

We start with a Type IIB-like model\foot{By this we mean simply
a closed string model with left- and right-moving world-sheet
supersymmetry and the same GSO projection for left- and right-movers.} 
with target space $\S^1$.
We permit unoriented worldsheets, but only in a way that is correlated
with the map to $\S^1$.  In fact, we require that the orientation
of the worldsheet $\Sigma$ is 
reversed in going around a loop $\gamma\in \Sigma$
if and only if $\gamma$ wraps around $\S^1$ an odd number of times.
A succinct way to say this is as follows.  We let $\alpha$ be the generator
of $H^1(\S^1,\Z_2)$.  We permit only maps $\Phi:\Sigma\to\S^1$ such
that $\Phi^*(\alpha)=w_1(\Sigma)$.   Here $w_1(\Sigma)$ (the first
Stieffel-Whitney class of $\Sigma$)
is the one-dimensional cohomology class that assigns the value $-1$
to a closed loop on $\Sigma$ if the orientation of $\Sigma$ is reversed
in going around $\Sigma$ and $+1$ otherwise.

In this model, the ${\bf {RP}}^2$ contribution to the vacuum amplitude
vanishes, for the following reason.  ${\bf {RP}}^2$ can
be obtained from a disc $D$ by imposing a certain equivalence relation
on the boundary of $D$.  If the boundary, which is of course  a circle,
is parametrized by an angular variable $\psi$, with $0\leq \psi\leq 2\pi$,
then one can take the equivalence relation to be $\psi\to\psi+\pi$.
Since the orientation of $\bf{ RP}^2$ is reversed in going from
$\psi=0$ to $\psi=\pi$, the restriction to maps $\Phi$ with
$\Phi^*(\alpha)=w_1(\Sigma)$ means that we must consider only maps
$\Phi$ such that, when $\psi$ is increased from $0$ to $\pi$, one wraps
$2n+1$ times around the target $\S^1$, for some integer $n$.
Therefore, the boundary of $D$ -- the full region $0\leq \psi\leq 2\pi$ --
is wrapped by $\Phi$ a total of $4n+2$ times around the target $\S^1$.
In particular, the wrapping number is always nonzero, so such a $\Phi$
cannot be extended over all of $D$, that is over ${\bf {RP}}^2$.  This
confirms that the ${\bf RP}^2$ contribution to the vacuum amplitude
vanishes.

To what physical processes can ${\bf RP}^2$ contribute?  Clearly,
it can contribute only to the expectation value of a product of vertex
operators for physical states whose total winding number is of the form
$4n+2$ for some integer $n$.  We will see momentarily that all physical
states in this theory have even winding number.  So if we call the winding
number $w$, then the free closed string sector has a $\Z_2$
symmetry that multiplies a state of winding $w$ by
$(-1)^{w/2}$.  The fact that all amplitudes on ${\bf {RP}}^2$ violate
conservation of winding number by $4n+2$ units for some integer $n$
means that the ${\bf RP}^2$ contribution is odd under $(-1)^{w/2}$.
We compare this to the fact that, as explained at the end of the last
subsection, the ${\bf RP}^2$ amplitude of the $(1,1)$ orientifold
is odd under $(-1)^p$.   We are led to suspect that there is a duality
between these theories that maps $w/2$ to $p$; as in our previous
experience, the factor of 2 corresponds to an unusual factor of 2 in
the radius of the $T$-dual to the $(1,1)$ orientifold.
(In the $(1,1)$ orientifold and its dual, the quantities that we have
called $p$ and $w/2$ are only defined mod 2, because of the existence
of unoriented worldsheets in one case and an orientifold projection in the
other.) 

Now let us analyze the spectrum of this model.  For this we work
in a Hamiltonian description.  We take $\Sigma$ to be the product
of a circle $C$, on which we quantize our strings, times a 
``time'' direction, parametrized by $\tau$, $0\leq \tau\leq T$ for some $T$.
In gluing $\tau=T$ to $\tau=0$, one may either reverse or not reverse
the orientation of $C$, so that the worldsheet is 
 either a torus or a Klein bottle.
However, in going around $C$, the orientation of $\Sigma$ is not reversed.

Because of that last fact, the restriction $\Phi^*(\alpha)=w_1(\Sigma)$
means that $C$ must wrap an {\it even} number of times around the
target space ${\bf S}^1$.  Thus, the winding number $w$ is even,
as was asserted earlier.

Let $R$ be the radius of the target ${\bf S}^1$.  The fact that only
even windings are allowed in this model suggests that its spectrum
may have a simple comparison to a more standard string model on a circle
of radius $2R$.  We will now see this by examining the
windings in the $\tau$ direction to determine the spectrum of momenta.

For a conventional oriented closed string mapped to a circle of radius $R$,
the sum over momentum states comes from the following factors in the
worldsheet
path integral.  There is a factor of $R$ for the center of mass of the
worldsheet, and also a sum over windings in the time direction.
Together, after a Poisson summation, these factors
give a sum over momentum states of the string of the form
 \eqn\uj{\sum_{n\in \Z}q^{n^2/R^2},} where $q$ depends on the
length of the torus (or Klein bottle) in the time direction.

Now in the present problem, because $\Phi^*(\alpha)=w_1(\Sigma)$,
the sum over windings in the time direction is modified.  Only
even windings are allowed if the worldsheet is a torus, and only
odd windings are allowed if the worldsheet is a Klein bottle.

Let us first think about the torus contribution.  Having only
even windings has roughly the same effect as doubling the radius of
the target space, which would double the basic allowed unit of winding.
But a factor of 2 is missing, since in our problem one gets a factor
of $R$ from the string center of mass position, while a conventional closed
string on a circle of radius  $2R$ would get a factor of 
$2R$ from the center of mass position.
So the contribution of the even time windings (that is the even
windings in the time direction) can be obtained from \uj\ by
replacing $R$ by $2R$ and dividing by 2, and so is
\eqn\juj{\half\sum_{n\in \Z}q^{n^2/(2R)^2}.}

The sum over odd time windings is of course  the difference
between the sum \uj\ over all windings and the sum \juj\ over even
windings.
  In our problem, the sum over  the odd windings
is accompanied by an operation $\Omega$ that reverses the worldsheet
orientation, so we represent this contribution symbolically
by $\Omega$ times the difference between \uj\ and \juj\ or
\eqn\ujuj{{\Omega\over 2}\sum_{n\in 2\Z}q^{n^2/(2R)^2}-{\Omega\over 2}
\sum_{n\in 2\Z+1}q^{n^2/(2R)^2}.}

The sum of \juj\ and \ujuj\ can be written
\eqn\pujuj{{1+\Omega\over 2}\sum_{n\in 2\Z}q^{n^2/(2R)^2}
+{1-\Omega\over 2}\sum_{n\in 2\Z+1}q^{n^2/(2R)^2}.}
Only by the minus sign in the last term does this differ from the
conventional momentum sum and orientation projection of a standard
Type I string on a circle of radius $2R$.  That sum would read,
in the same notation,
\eqn\upujuj{{1+\Omega\over 2}\sum_{n\in \Z}q^{n^2/(2R)^2}.}

Comparing the last two formulas, we see that the present model,
on a circle of radius $R$, has for even momentum precisely the
same states as a conventional Type I superstring on a circle
of radius $2R$.  For odd momentum, the model considered here
has an orientation projection $(1-\Omega)/2$ while the standard
Type I superstring has an orientation projection $(1+\Omega)/2$.
However, since $\Omega$ changes the sign of the momentum, the masses
and quantum numbers that one gets for states
of non-zero momentum, and therefore in particular for states of odd momentum,
are the same whether one projects onto $\Omega=1$ or onto $\Omega=-1$.
Consequently, the present model on a circle of radius $R$ has the
same spectrum as the ordinary Type I string on a circle of radius
$2R$, but the detailed  wavefunctions of physical states
are different, and consequently the interactions will be different.
As summarized at the end of section 6.1, this is the expected behavior for
the $T$-dual that we are seeking of
the $(1,1)$ orientifold on $\S^1/\Z_2$.

\bigskip\noindent{\it Systematic Perturbation Expansion Of The Orientifold}

To be more precise in comparing the model just described to the
$(1,1)$ orientifold on $\S^1/\Z_2$, we need a precise
recipe for systematic worldsheet perturbation theory on the 
orientifold.  We describe this for a general orientifold that
does not have an open string sector (one could also elaborate the
description to incorporate the open strings), and then specialize
to the case that the target space is $\R^9\times \S^1$.

We consider a spacetime $M$ with an involution $\tau_M$ (an involution
is simply a symmetry whose square is the identity).  The Riemann
surfaces in the theory will be orientable Riemann surfaces $\Sigma$ with
orientation-reversing freely-acting involution $\tau_\Sigma$.  
(One allows the case in which $\Sigma$ consists of two identical
components exchanged by $\Sigma$, as a result of which the quotient
$\Sigma/\tau_\Sigma$ is orientable but not endowed with an orientation,
and also the case in which  $\Sigma$ is connected and the quotient
$\Sigma/\tau_\Sigma$ is unorientable.)  The worldsheet path
integral of this theory will be taken over maps $\Phi:\Sigma\to M$
which commute with the $\tau$'s in the sense that $\Phi\circ \tau_\Sigma
=\tau_M\circ \Phi$.  (Such an equivariant formalism was used in 
\ref\ohorava{P. Horava, ``Equivariant Topological Sigma Models,''
Nucl. Phys. {\bf B418} (1994) 571, hep-th/9309124.}.)

The most standard worldsheet action of this theory would be the
usual Nambu-Goto action
\eqn\uggu{I={1\over 2}{1\over 4\pi\alpha'}\int_\Sigma\sqrt g g^{ab}\partial_a
X^I\partial_bX^JG_{IJ}.}
The factor of $\half$ is included so as to normalize
the action in the standard way 
for maps from $\Sigma/\tau_M$ rather than for maps from 
$\Sigma$.  The path integral of the theory will run over $\tau$-invariant
maps, with an integrand that we could take to be simply $e^{-I}$.
We want to generalize this, to obtain theories with orientifold
planes of different kinds.

To do this, we take an arbitrary  element $x\in H^1_{\bf \Z_2}(M,\Z_2)$,
the $\Z_2$-equivariant cohomology of $M$ with $\Z_2$ coefficients; here
$\Z_2$ is the group generated by $\tau_M$.
Then $\Phi^*(x)$ is an element of $H^1_{{\bf Z}_2}(\Sigma,\Z_2)$, where
now $\Z_2$ is generated by $\tau_\Sigma$.
Since $\Z_2$ acts freely on $\Sigma$, this is the same as $H^1(\Sigma/\Z_2,
\Z_2)$.
The obstruction $w_1(\Sigma/\Z_2)$ to orientation of $\Sigma/\tau_\Sigma$
also takes values in $H^1(\Sigma/\Z_2,\Z_2)$, and there is a well-defined
mod 2 pairing $(w_1(\Sigma/\Z_2),\Phi^*(x))$.  We can incorporate
this pairing in the definition of the worldsheet Lagrangian and take
the integrand to be
\eqn\jurry{e^{-I}(-1)^{(w_1(\Sigma/\Z_2),\Phi^*(x))}.}

The advantage of this somewhat abstract definition is that, because
of its cohomological nature, it is manifest that all factorization
or cut and paste requirements of string theory are satisfied.  
Hence one does get a systematic string perturbation theory for any
$x$.

We will now specialize to the case that $M$ is $\R^9\times \S^1$,
with $\S^1$ parametrized by an angle $\theta$,
and $\tau_M$ is $\theta\to -\theta$.
The goal is to show that for suitable $x$, we do get the desired
$(1,1)$ orientifold with one fixed point of type ${\cal O}^-$ and
one of type ${\cal O}^+$.  
To do this, we take $x$ to be the 
cohomology class dual to a $\Z_2$-invariant cycle in $M=\R^9\times \S^1$
consisting of one of the orientifold planes, say the one at 
$\theta=0$.  Now we consider the case that $\Sigma=\S^2$ and
$\Sigma'=\Sigma/\Z_2$ is
${\bf{RP}}^2$.  If $\Phi$ is a constant map from $\Sigma$ to the
orientifold plane at $\theta=\pi$ (which induces a constant map
from ${\bf RP}^2$ to that orientifold plane), then $\Phi^*(x)=0$, because
the support of $x$ is disjoint from $\theta=\pi$.  If, however, $\Phi$
is  a constant map from ${\bf RP}^2$ to the orientifold
plane at $\theta=0$, then the normal bundle to the orientifold plane
pulls back to the ``orientation bundle'' (the determinant of the tangent
bundle) of ${\bf{RP}}^2$.  In this case, $\Phi^*(x)=w_1({\bf{ RP}}^2)$
and is in particular non-zero.  Also, $(w_1({\bf {RP}}^2),\Phi^*(x))
=w_1({\bf{RP}}^2)^2=1$ modulo 2.  So  the sign factor in \jurry\
is $-1$ for this map, confirming that the $\theta=0$ orientifold
is of opposite type from the $\theta=\pi$ orientifold.

I will not attempt to give in this paper a formal proof that
the orientifold with $x$ as in the last paragraph is $T$-dual to the 
model of section 6.2.  However, $T$-duality brings about a kind
of Fourier transform, and would be expected to map a sign
factor $(-1)^{(w_1(\Sigma/\Z_2),\Phi^*(x))}$ to a delta function
setting $w_1(\Sigma/\Z_2)$ equal to $\Phi^*(x)$. This delta
function was used in the definition of the model of section 6.2.

\bigskip\noindent{\it Answer To The Original Question}

At this point, we can answer the original question, which was to
describe a $T$-dual of the $(2,2)$ orientifold on $\T^2/\Z_2$.
I claim that this is obtained simply by compactifying on a circle
the model  of section 6.2 which is $T$-dual to the $(1,1)$
orientifold on $\S^1/\Z_2$.

The $(2,2)$ orientifold on $\T^2/\Z_2$ can be described as follows
in the language of the present discussion.  Take spacetime
to be $M=\R^8\times \T^2$.  Let $\tau_M$
be the usual involution that acts as $-1$ on $\T^2$.  Now consider
an orientifold on $M/\tau_M$ with the following choice of $x$.
If $\T^2=\S^1\times \S^1$, and $\theta$ is an angular parameter
on the first $\S^1$, let $x$ be the Poincar\'e dual to the $\tau$-invariant
hypersurface $\theta=0$.
Consider the orientifold model with this choice of $x$
and the action as in \jurry.  An analysis as in the last subsection
shows that the two orientifold planes at $\theta=0$ are of one type
and the two at $\theta=\pi$ are of the other type, so this model is the
$(2,2)$ orientifold.

On the other hand, $T$-duality will turn $\T^2/\Z_2$ into a dual
$\T^2$, and will  Fourier transform the sign
factor $(-1)^{(w_1(\Sigma/\Z_2),\Phi^*(x))}$ to 
a delta function setting $w_1(\Sigma/\Z_2)$ equal to $\Phi^*(x)$.
The model on $\T^2=\S^1\times \S^1$ with this delta function
is simply the compactification on $\S^1$ (the $\S^1$ in question
being the second factor of $\T^2=\S^1\times \S^1$) of the model
that we introduced and analyzed in section 6.2 and
 found to be $T$-dual to the $(1,1)$ orientifold on
$\S^1/\Z_2$.

\bigskip\noindent{\it Systematic Perturbative Description Of The $(3,1)$
Orientifold}

We have given a systematic and manifestly factorized description
of the worldsheet perturbation expansion for the $(1,1)$ orientifold
in nine dimensions, and the $(2,2)$ orientifold in eight dimensions.
At this point, one might like to describe the eight-dimensional
$(3,1)$ orientifold, which has been of course the main subject of
the present paper, in an analogous way.  We will do so,
at least for the closed string sector.

For this, we    return first to the general formulation
of orientifold perturbation theory, in terms of maps from
a Riemann surface $\Sigma$ with a free, orientation-reversing involution
$\tau_\Sigma$ to a spacetime $M$ that is equipped with an involution
$\tau_M$.  The path integral is over maps $\Phi:\Sigma\to M$ such
that $\tau_\Sigma\circ\Phi=\Phi\circ\tau_M$.  Suppose that
we are given an element $y\in H^2_{\Z_2}(M,\Z_2)$.  The
pullback $\Phi^*(y)$ is an element of $H^2_{\Z_2}(\Sigma,\Z_2)$, which
since $\tau_\Sigma$ acts freely on $\Sigma$ is the same as
$H^2(\Sigma/\Z_2,\Z_2)$.  So there is a well-defined sign factor
\eqn\jklp{(-1)^{\int_{\Sigma/\Z_2}\Phi^*(y)}.}
Because of its cohomological nature, its inclusion in the path
integrand is manifestly compatible with all factorization requirements
of string theory.

Let us implement this procedure for the case that $M=\R^8\times\tilde\T^2$,
with $\tilde \T^2$ the ``dual torus'' parametrized by angles
$\theta,\psi$.  We take $\tau_M$ to be the involution
$\theta\to-\theta$, $\psi\to-\psi$, with the usual four orientifold
fixed points at which $\theta $ and $\psi$ are both 0 or $\pi$.
We take $y$ to be the Poincar\'e dual to the $\Z_2$-invariant
cycle $\theta=\psi=0$.
I claim that in this case, inclusion of the factor \jklp\ in the path
integrand has the effect of reversing the ``type'' of the orientifold
plane at $\theta=\psi=0$ without affecting the others, so that
it generates the closed
string sector of the $(3,1)$ orientifold from that of
the standard $(4,0)$ orientifold.

For this it suffices to show that, if $\Sigma=\S^2$, $\Sigma/\tau_\Sigma
={{\bf {RP}}}^2$, with $\Phi$ a constant map to an orientifold fixed
point, then the sign factor in \jklp\ is $-1$ if $\theta=\psi=0$
and otherwise $+1$.  The sign factor is $+1$ for $(\theta,\psi)
\not= (0,0)$  since the orientifold fixed points away from
the origin are disjoint from the support of $y$.  On the other hand,
for $\Phi$ a constant map to $\theta=\psi=0$, the pullback of the
tangent bundle of $\tilde \T^2$ to ${\bf {RP}}^2$
is the tangent bundle of ${\bf {RP}}^2$, so in this case the
sign factor in \jklp\ reduces to $(-1)^{w_2({\bf {RP}}^2)}=-1$.

For $\tilde \T^1/\Z_2$ and $\tilde \T^2/\Z_2$ orientifolds,
the cohomological formulas we have described enable one to develop
a systematic worldsheet perturbation expansion with an arbitrary
labeling of fixed points as being of type ${\cal O}^+$ or ${\cal O}^-$.
(In some cases one will run into trouble because of noncancellation
of brane charges, but the formal rules make sense and are compatible
with factorization.)  For $\tilde \T^n/\Z_2$ with $n>2$, it does not
appear that an arbitrary assignment of the ``type'' of the orientifold
planes will give a theory that can be described in a systematic
string perturbation expansion.  For restrictions on the orientifold
configuration in an analogous situation, see \ref\otherpol{J. Polchinski,
``Tensors From K3 Orientifolds,'' Phys. Rev. {\bf D55} (1997)
6423, hep-th/9606165.}.

\appendix{I}{A Note On Four-Dimensional Gauge Theories}

In this appendix we suggest a resolution to
a longstanding puzzle concerning the
computation of the supersymmetric index $\Tr\,(-1)^F$ in supersymmetric
gauge theories in four dimensions.  As in much of  the
body of the paper, the key point will be to understand certain facts
about the components of the moduli space of flat $\Spin(n)$ connections
on a torus.

Consider four dimensional supersymmetric Yang-Mills
theory, with a connected, simple, and simply-connected
\foot{As in section 7 of \ref\indexw{E. Witten, ``Constraints
on Supersymmetry Breaking,'' Nucl. Phys. {\bf B202} (1982) 253.},
one can also consider the case that $G$ is not simply-connected
(and/or not connected), in which case one will encounter topologically
non-trivial $G$ bundles on $\T^3$.  However, to illustrate the
essential issues we wish to consider here, it suffices to focus
on the case that $G$ is connected and simply-connected, in which
case the bundle is topologically trivial.} gauge
group $G$
and no chiral superfields.  This model has a discrete chiral
symmetry group $\Z_{2h}$, where $h$ is the dual Coxeter number.

It is believed that this model undergoes
spontaneous chiral symmetry breaking, with $\Z_{2h}$ spontaneously
broken to $\Z_2$ (its maximal subgroup that permits gluino bare masses).
This results in the existence of $h$ distinct vacua, each of which
is believed to have a mass gap (and confinement).  If one formulates
the theory on  $\R^1\times \T^3$, with $\R^1$ being the ``time direction,''
and $\T^3$ being a spatial three-torus, then one expects a vacuum
with a mass gap to contribute $+1$ to $\Tr\,(-1)^F$.   (The contribution
of a vacuum without a mass gap would not necessarily be $+1$.) One therefore
expects $\Tr\,(-1)^F=h$.

\def\M{{\cal M}}
On the other hand, an explicit computation of $\Tr\,(-1)^F$
was made in section 8 of
\indexw\ by actually counting, for weak coupling, the supersymmetric states on
a torus.  This was done by first finding the classical moduli space
${\cal M}$ of zero energy states  on $\T^3$,
and then performing a Born-Oppenheimer quantization.  The classical
states of zero energy are given by representations of the fundamental
group of $\T^3$ in $G$, or in other words by a choice (up to conjugation)
of three commuting elements
$U_1,U_2$, and $U_3$ of $G$.  
It was argued that by quantizing the component of $\M$ that contains
$U_1=U_2=U_3=1$, one gets a contribution $r+1$ to $\Tr\,(-1)^F$,
where $r$ is the rank of $G$.

One is thus led to expect that $h=r+1$. This is in fact so for $SU(n)$
(with $h=n$, $r=n-1$),
 and it is also true for $Sp(n)$
(with $h=n+1$, $r=n$).  The puzzle is that it is not so for $\Spin(n)$
with $n\geq 7$ that $h=r+1$. (The $\Spin(n)$ groups with $n=3,5$, or 6
do work as they are equivalent to $SU(2)$, $Sp(2)$, and $SU(4)$.  $\Spin(4)$
is not simple so the above discussion does not precisely apply to it, but
since $\Spin(4)=SU(2)\times SU(2)$, a slightly corrected version of the
formula does work for $\Spin(4)$.) 
  On the contrary, for $\Spin(2k)$ with
$k\geq 3$ one has $h=2k-2$, $r=k$, and for $\Spin(2k+1)$ with $k\geq 2$
one has $h=2k-1$, $r=k$.

The error in \indexw\ was to assume that $\M$ is connected and to
evaluate the contribution only of the component of $\M$ containing
$U_1=U_2=U_3=1$, which we will call the trivial flat connection.  
It is true for $SU(n)$ that $\M$ is connected.
In fact, any family of commuting elements of $SU(n)$, such as 
$U_1$, $U_2$, and $U_3$, can be simultaneously
diagonalized, or in other words conjugated to a maximal torus.
As the maximal torus is connected and abelian, this means that $\M$ is
connected.
   Likewise any family of commuting elements of $Sp(n)$
can be conjugated to a maximal torus (a proof of this by induction on
the number of commuting group elements
uses the fact that any one element of $Sp(n)$ has the following properties:
{(\it i)} it can be conjugated to a maximal torus; {(\it ii)} its commutant
is a product of unitary and symplectic groups).  So
$\M$ is connected for $Sp(n)$.  $\M$ is likewise connected for $\Spin(n)$
with $n\leq 6$ because of equivalences mentioned in the last paragraph.
But this fails for $\Spin(n)$ with $n\geq 7$, as we have seen in section
3.3, and this is the reason
that for those groups one should not expect $h$ to equal $r+1$.

The general formula for arbitrary simple, connected, and simply-connected
$G$ can, however, be worked out by extending the ideas in \indexw.
Let $\M_i$ be the connected components of $\M$.  Let $G_i$ be a maximal
unbroken subgroup of a flat connection determined by a point in $\M_i$.
Then, as in \schwei\ and our discussion in section 2 for the $\T^2$ case,
$\M_i$ is up to a finite cover (which will not affect the following
discussion) the same as the moduli space of triples of
commuting elements $V_1,V_2,V_3$ of $G_i$ that are continuously
connected to $V_1=V_2=V_3=1$.  
The contribution of $\M_i$ to $\Tr\,(-1)^F$ can be computed by the same
computation as in \indexw, but with $G$ replaced by $G_i$.
Hence, if $r_i$ is the rank of $G_i$
(there may be several possible $G_i$, as in the analogous case
treated in section 2, but they all have the same rank), the contribution
of $\M_i$ to $\Tr\,(-1)^F$ is $r_i+1$.\foot{This really assumes
that the $G_i$ are simple  (which will be the case for the examples
we consider) since the computation in \indexw\ was for simple groups.
Also, the computation in \indexw\ was for connected gauge groups,
while in examples below the $G_i$ will not always be connected.
In fact, we will meet an example in which one of the $G_i$ is $O(k)$
(or rather its double cover ${\rm Pin}(k)$) rather than $SO(k)$.  By
a reexamination of the argument in \indexw, one can see that the formula
$r+1$ gives the correct result for $O(k)$ for all $k$,
even in the exceptional case
$k=2$.  (This would not be so for the non-simple group $SO(2)$, for
which $\Tr\,(-1)^F=0$, so in the discussion below getting the right
value of $\Tr\,(-1)^F$  for $\Spin(9)$ depends on the fact that the
relevant centralizer is $O(2)$ instead of $SO(2)$.)}

  Summing over all components,
the formula saying that the analysis in terms of physical vacua and
chiral symmetry breaking should agree with the explicit weak coupling
analyis is
\eqn\hogov{h=\sum_i(r_i+1)}
if the $G_i$ are all simple.

Let us verify this formula for $\Spin(n)$.  We start with $\Spin(7)$, which is
the first problematical case.  Apart from a component $\M_1$ consisting of flat
connections that are continuously connected to the identity, the moduli
space contains a component $\M_2$ that is an isolated point. The flat
connection corresponding to the unique point in $\M_2$ can be described,
up to conjugation, by commuting holonmies $U_1,U_2,U_3$, which are
diagonal elements with eigenvalues $(U_1,U_2,U_3) =(\pm 1,\pm 1, \pm 1)$,
with each of the seven combinations of signs other than $(1,1,1)$ appearing
with multiplicity one.  (As was sketched in a footnote in section 3.3,
this bundle has $w_1=w_2=0$, and so is a $\Spin(7)$ bundle and in
particular is topologically trivial.)
This flat connection admits no deformations; its centralizer (the unbroken
subgroup of $\Spin(7)$) is a finite
group, of rank 0.  Meanwhile, the centralizer of $\M_1$ is, of course,
$\Spin(7)$, of rank 3.
That $\M_1$ and $\M_2$ are the only components of $\M$ can be proved
using the $D$-brane description of $\Spin(7)$ flat connections;
we postpone this argument until the end of the present appendix.
So the identity \hogov\ becomes $5=(3+1)+(0+1)$, compatible with the
standard conjectures about the dynamics of the $\Spin(7)$ theory.

It is straightforward to generalize this to $\Spin(n)$ with $n>7$.  $\M$
has a component $\M_1$ that contains the trivial flat connection; the rank
of its centralizer is that of $\Spin(n)$.  There is as explained below
precisely one more component,
which parametrizes a family of flat connections of which one example is
given by the
commuting holonomies $V_i=U_i\oplus {\bf 1}_{n-7}$, where $U_i$
are the $\Spin(7)$ matrices of the last paragraph and ${\bf 1}_{n-7}$ is
an $n-7$-dimensional identity matrix.  The rank of the unbroken subgroup
is that of $\Spin(n-7)$.\foot{The unbroken subgroup itself is not
$\Spin(n-7)$ but ${\rm Pin}(n-7)$ (an extension of $\Spin(n-7)$ to allow
orientation-reversing symmetries) times some $\Z_2$'s.  That one gets
${\rm Pin}(n-7)$ instead of $\Spin(n-7)$ is important for $n=9$, for
a reason explained in the last footnote.}  
The sum of the rank of $\Spin(n)$ and that
of $\Spin(n-7)$ is $r_1+r_2=n-4$, while for $\Spin(n)$ one has
$h=n-2$.  So we get the expected identity $h=(r_1+1)+(r_2+1)$, compatible
with the standard conjectures concerning the dynamics of the $\Spin(n)$
theory.

One expects \hogov\ to hold also for the exceptional Lie groups (perhaps
with some modifications if the $G_i$ are not simple),
but an efficient verification of this really requires a more powerful
method of computation.

It remains to verify that for $\Spin(n)$ with $n\geq 7$, there are precisely
the two components of the moduli space of flat connections claimed above.
First we consider $n=7$.
To describe a $\Spin(7)$ flat connection, one needs a $\Z_2$-invariant
configuration of seven $D$-branes on $\tilde \T^3$, the covering space of
an orientifold $\tilde \T^3/\Z_2$.  One of the
orientifold planes, call it ${\cal O}_0$, corresponds to eigenvalues
$(1,1,1)$ of $U_1,U_2,U_3$, and the others, call them
${\cal O}_\alpha$, $\alpha=1,\dots,7$, to sequences $(\pm 1,\pm 1,\pm 1)$ with
at least one $-1$.  If there are fewer than seven orientifold planes
at which the number of  zerobranes is odd, 
then by using the fact that the moduli
space of $\Spin(n)$ flat connections on $\T^3$ is connected for $n<7$
(or simply by a direct computation),
one shows that either the flat bundle has non-zero $w_1$ or $w_2$ and
so is not a $\Spin(7)$ bundle and is topologically non-trivial (and
does not contribute to $\Tr\,(-1)^F$ with untwisted boundary conditions),
or it is continuously connected via flat connections
to $U_1=U_2=U_3=1$.  So we need only
consider the case that  seven of the eight orientifold planes contain 
a single $D$-brane each.  There is hence exactly one orientifold plane
that has no $D$-branes.  If it is not ${\cal O}_0$, then the bundle
has $w_1\not=0$ and is again not a $\Spin(7)$ bundle.  So the only
relevant
case is the case of precisely one $D$-brane at each of the ${\cal O}_\alpha$,
and this gives the flat $\Spin(7)$ connection with holonomies
$U_i$ that was described earlier.
For $\Spin(n)$ with $n>7$, one shows by the
same argument that the only component of $\Spin(n)$ flat bundles on $\T^3$
that is not continuously connected to the trivial flat bundle is obtained
as follows.  It $n$ is even, one places
 one $D$-brane at each orientifold plane and lets the
others wander in pairs. If $n$ is odd, one places one $D$-brane at each
of the ${\cal O}_\alpha$ and lets the others wander in pairs.  In particular,
for all $n\geq 7$ there is precisely one component of the moduli space
of flat $\Spin(n)$
connections apart from the component of the trivial flat connection.
An example of a flat connection in this component is obtained 
(whether $n$ is even or odd) by placing
one $D$-brane at each of the ${\cal O}_\alpha$ and $n-7$ at ${\cal O}_0$;
the corresponding holonomies are $V_i=U_i\oplus {\bf 1}_{n-7}$ with
$U_i$ as above.

\bigskip\noindent
{\it Acknowledgments}

I would like to thank A. Borel,
J. H. Conway, P. Mayr, J. Morgan, Y. Oz, Jaemo Park, N. Seiberg, A. Uranga,
and S. Weinberg for helpful suggestions and comments.  This work
was supported in part by NSF Grant PHY-9513835.

\listrefs
\end